\documentclass[conference]{IEEEtran}
\IEEEoverridecommandlockouts
\usepackage{cite}
\usepackage{amsmath,amssymb,amsfonts}
\usepackage{algorithmic}
\usepackage{graphicx}
\usepackage{textcomp}
\usepackage{xcolor}
\usepackage{tabularx}
\usepackage{graphics}
\usepackage{subfigure}
\usepackage{makecell}
\usepackage{multirow}
\usepackage{multicol}
\usepackage{array}
\usepackage{threeparttable}
\usepackage[most]{tcolorbox}
\usepackage{url}

\def\BibTeX{{\rm B\kern-.05em{\sc i\kern-.025em b}\kern-.08em
    T\kern-.1667em\lower.7ex\hbox{E}\kern-.125emX}}
    
\begin{document}

\title{Revisiting Machine Learning based Test Case Prioritization for Continuous Integration}

\author{
\IEEEauthorblockN{1\textsuperscript{st} Yifan Zhao}
\IEEEauthorblockA{\textit{Key Laboratory of High Confidence} \\
\textit{Software Technologies (Peking } \\
\textit{University), Ministry of Education} \\
Beijing, China \\
zhaoyifan@stu.pku.edu.cn}
\and
\IEEEauthorblockN{2\textsuperscript{nd} Dan Hao$^{\ast}$ \thanks{$^{\ast}$Dan Hao is the corresponding author.}}
\IEEEauthorblockA{\textit{Key Laboratory of High Confidence} \\
\textit{Software Technologies (Peking } \\
\textit{University), Ministry of Education} \\
Beijing, China \\
haodan@pku.edu.cn}
\and
\IEEEauthorblockN{3\textsuperscript{rd} Lu Zhang}
\IEEEauthorblockA{\textit{Key Laboratory of High Confidence} \\
\textit{Software Technologies (Peking } \\
\textit{University), Ministry of Education} \\
Beijing, China \\
zhanglucs@pku.edu.cn}
}

\maketitle

\begin{abstract}
To alleviate the cost of regression testing in continuous integration (CI), a large number of machine learning-based (ML-based) test case prioritization techniques have been proposed. However, it is yet unknown how they perform under the same experimental setup, because they are evaluated on different datasets with different metrics. To bridge this gap, we conduct the first comprehensive study on these ML-based techniques in this paper. We investigate the performance of 11 representative ML-based prioritization techniques for CI on 11 open-source subjects and obtain a series of findings. For example, the performance of the techniques changes across CI cycles, mainly resulting from the changing amount of training data, instead of code evolution and test removal/addition. Based on the findings, we give some actionable suggestions on enhancing the effectiveness of ML-based techniques, e.g., pretraining a prioritization technique with cross-subject data to get it thoroughly trained and then finetuning it with within-subject data dramatically improves its performance. In particular, the pretrained MART achieves state-of-the-art performance, producing the optimal sequence on 80\% subjects, while the existing best technique, the original MART, only produces the optimal sequence on 50\% subjects.
\end{abstract}

\begin{IEEEkeywords}
test prioritization, machine learning, continuous integration
\end{IEEEkeywords}

% \vspace{-0.3cm}
\section{Introduction}\label{sec:introduction}
Continuous Integration (CI) is a software development practice. It is widely used in industry, making software release more rapid and reliable~\cite{hilton2016usage}. 
In the CI environment, developers continuously commit code to modify software functionalities.

To guarantee the quality of committed code, regression testing~\cite{ji2022automated} tends to be conducted in each CI cycle, which is widely recognized to be time-consuming~\cite{rothermel2001prioritizing} and may hamper rapid software release. 
% A large number of test case prioritization (TCP) techniques have been proposed in the literature to alleviate the cost problem, but most of them target the general regression testing process, rather than regression testing in CI. Moreover, these techniques cannot be directly applied to the latter scenario, because a large amount of modification on code and test cases occurs during frequent CI, which most existing TCP techniques cannot appropriately deal with~\cite{DBLP:conf/icse/LuLCZHZ016,DBLP:conf/sigsoft/WangNT17}. 
Many test case prioritization (TCP) techniques have been proposed to alleviate the cost of regression testing. Still, most of them target the general regression testing process. They cannot be directly applied to regression testing in CI, because a large amount of modification on code and tests occurs during frequent CI, which most existing TCP techniques cannot appropriately deal with~\cite{DBLP:conf/icse/LuLCZHZ016,DBLP:conf/sigsoft/WangNT17}.

To address the specific problem of TCP in CI, heuristic-based and machine learning-based (ML-based) techniques are proposed. Some researchers give the first attempt by scheduling tests based on heuristics (e.g., time since the last test failed~\cite{elbaum2014techniques} or test diversity~\cite{haghighatkhah2018test}).
Other researchers harness the power of machine learning by using a large amount of historical data in CI, and propose numerous ML-based TCP techniques which have been demonstrated to be promising. These ML-based techniques build neural models to predict the optimal sequence of tests instead of human-defined strategies. In particular, these ML-based techniques can be categorized into supervised learning-based (SL-based)~\cite{busjaeger2016learning,bertolino2020learning,sharif2021deeporder} and reinforcement learning-based (RL-based) techniques~\cite{spieker2017reinforcement,bagherzadeh2021reinforcement,do2020multi,yang2020systematic}. 
In training cycles, an SL-based technique trains a classification model based on tests and their labels. In testing cycles, the model is used to predict priority values for tests. That is, once the model completes training, it uses a fixed strategy to prioritize testing instances.
In contrast, the RL-based technique continuously adjusts its prioritization strategy. 
In each cycle, it is first tested (i.e., prioritizing the test suite) and then trained based on the prioritization feedback. In other words, it incrementally trains on coming data.
Although ML-based techniques are widely studied in TCP for CI as reflected by a series of publications, it remains unknown how the SL-based and RL-based techniques perform under the same experimental setup, because they are evaluated on different datasets with different input and metrics~\cite{bagherzadeh2021reinforcement,bertolino2020learning,sharif2021deeporder,busjaeger2016learning,do2020multi}. Note that a few prior works~\cite{yaraghi2022scalable, elsner2021empirically} compare several SL-based techniques through empirical studies, but they ignore recently proposed RL-based techniques. Therefore, a developer will find it hard to choose an effective TCP technique to use in CI, and a researcher will find it hard to conduct more profound research without a commonly-agreed setup.

To investigate the performance of ML-based TCP in CI, in this paper we conduct a comprehensive experimental study on 11 highly-starred open-source projects from GitHub. Through the systematic analysis, we find that the metrics (i.e., AFPD, NAPFD, and NRPA) used by the existing work have problems in fair comparison and discernment. Therefore, we propose the rectified APFD (rAPFD), with which we evaluate 11 representative ML-based TCP techniques (including six SL-based techniques and five RL-based techniques), in terms of their effectiveness, efficiency and applicability in CI. 

Our study reveals a series of findings. \textbf{First, data imbalance is a common phenomenon in TCP for CI.} \textbf{On more-failure and less-failure subjects} (i.e., subjects with higher or lower ratio of failing tests to all tests), \textbf{the performance of ML-based TCP techniques is different, which is caused by data imbalance. Second, the performance of ML-based TCP techniques changes across CI cycles, mainly caused by the changing amount of training data instead of code evolution and test removal/addition.} More training data may 
improve the effectiveness of TCP techniques. Third, RL-based techniques generally have a much longer training time than SL-based techniques. \textbf{Fourth, the RL-based technique PPO1-LI is inapplicable to CI in some subjects, whereas the other techniques are generally applicable to the CI context. The SL-based technique MART helps developers save more than 95\% testing time in most subjects.} Based on the findings, we give some actionable suggestions that can boost existing techniques, supported by our additional experiments. \textbf{First, techniques robust to data imbalance, like ensemble learning or cost-sensitive techniques, are promising in this scenario.} Besides, over-sampling approaches like SMOTE can be applied to alleviate the imbalance problem and boost existing techniques. \textbf{Second, cross-subject data can be used to pretrain TCP techniques to get them thoroughly trained and improve their effectiveness.} It is also worth noting that we pretrain MART using cross-subject data and greatly enhance its effectiveness: The pretrained MART achieves state-of-the-art performance, producing the optimal sequence on 80\% subjects, while the original MART, which is reported to be the best technique~\cite{bertolino2020learning,yaraghi2022scalable}, produces the optimal sequence on only 50\% subjects. Based on the comparison of test duration and the prediction time of techniques, we also suggest that \textbf{test duration should be estimated to assess the necessity of applying any TCP technique and the overhead of TCP techniques deserves more attention.} Overall, this work advances our understanding of ML-based TCP techniques in CI and provides valuable insights for improving their effectiveness.

This paper presents three key contributions.
First, it offers the first comprehensive study on ML-based TCP techniques in CI. The study analyzes the impact of data imbalance, the quantity of training data, and code changes on the effectiveness of prioritization techniques. It also investigates the efficiency and applicability of the techniques. 
Second, it provides actionable suggestions to enhance existing ML-based TCP, based on the study's findings.
    % : (1) Ensemble learning, cost-sensitive techniques are promising in this scenario. Pre-processing approaches alleviating data imbalance can release the power of existing TCP techniques. (2) Cross-subject data can be used to pretrain TCP techniques to get them fully trained and greatly enhance their effectiveness. As a support for the second suggestion, the pretrained MART achieves state-of-the-art performance.
Third, a replication package is available at \textbf{\url{https://github.com/yifan-CodeDir/ml-citcp}}.
% \vspace{-0.3cm}
% \begin{itemize}
%     \item \textbf{The first comprehensive study} on ML-based TCP techniques in CI. We analyze how data imbalance, the amount of training data, and code change affect techniques' prioritization effectiveness. We also investigate the efficiency and applicability of the techniques. 
%     \item \textbf{Actionable suggestions} to boost existing ML-based TCP.
%     % : (1) Ensemble learning, cost-sensitive techniques are promising in this scenario. Pre-processing approaches alleviating data imbalance can release the power of existing TCP techniques. (2) Cross-subject data can be used to pretrain TCP techniques to get them fully trained and greatly enhance their effectiveness. As a support for the second suggestion, the pretrained MART achieves state-of-the-art performance.
%     \item A replication package available at \newline \textbf{\url{https://github.com/Anonymous6346/ml-citcp}}.
% \end{itemize}

\section{Background and related work}
% \dan{shorten the following description}
Continuous Integration automates the integration of code changes from multiple contributors and has been widely adopted by companies like Google, Facebook, Microsoft~\cite{liang2018redefining}. In CI environment, developers copy the software project to the local machine and introduce new features, repair bugs, and change existing test cases. The code changes are submitted and integrated into the ``mainline'' after local build and test, and the project is then rebuilt and tested, which completes a CI cycle. Through CI, developers can easily track the progress of development and reduce the risk of accumulating bugs.

%TCP was firstly proposed by Wong et al.~\cite{wong1997study}. Given a test suite $T$, the set of its permutations on tests $PT$, an objective function from $PT$ to a real number $f$, TCP aims to find $T^{\prime} \in PT$ such that $(\forall T^{\prime\prime})(T^{\prime\prime} \in PT)(T^{\prime\prime} \neq T^{\prime}) \left[f(T^{\prime}) \geq f(T^{\prime\prime})\right]$, where the function $f$ measures the value of permutations. 
TCP was first proposed in regression testing~\cite{wong1997study} to schedule the execution order of test cases for some goals, e.g., revealing faults as early as possible. 
A large amount of research contributes to the TCP domain, resulting in many surveys~\cite{Yooh12survey,CatalM13survey,rothermel2001prioritizing,lou2019survey}~ and applications in industry~\cite{machalica2019predictive,anderson2014improving,herzig2015art}. %The existing work can be classified into experimental studies and prioritization techniques. Apart from the studies on the performance of existing TCP techniques by considering various coverage criteria~\cite{henard2016comparing}, metrics~\cite{rothermel1999test,qu2007combinatorial}, and time constraints~\cite{zhang2009time,do2010effects,walcott2006timeaware}, researchers proposed various algorithms (e.g., greedy algorithms~\cite{li2010simulation,rothermel2001prioritizing}, search-based algorithms~\cite{li2007search}, integer linear programming based algorithms~\cite{zhang2009time}, ML-based algorithms~\cite{busjaeger2016learning,bertolino2020learning,spieker2017reinforcement}, and information-retrieval based algorithms~\cite{peng2020empirically}) for TCP. However, these general TCP techniques may not work in CI due to the following characteristics of CI. 
Although many TCP techniques~\cite{li2010simulation,rothermel2001prioritizing,li2007search,zhang2009time,busjaeger2016learning,bertolino2020learning,spieker2017reinforcement,peng2020empirically,hao2014unified,hao2013adaptive,lou2015mutation,li2021aga} have been proposed in the literature, these general TCP techniques may not work in CI due to the following characteristics of CI. 
First, frequent code commit and integration in CI may result in big code difference in various CI cycles. However, the existing TCP techniques usually schedule the tests based on the structural coverage (e.g., statement or method coverage), which may lose effectiveness in the current commit. Second, tests may be updated with the code in CI. The newly-added tests with no previous execution results and structural coverage information present a challenge for priority prediction. Third, TCP in CI has an extra time constraint (i.e., low overhead)~\cite{fowler2006continuous} since CI is proposed to speed up development. However, existing TCP techniques schedule the tests offline and thus they usually have no such constraint. To sum up, TCP in CI is different from TCP in existing scenarios. Since this paper focuses on ML-based TCP in CI, due to space limitation, in this section, we review only the work of TCP in CI.

% \subsection{Test case prioritization for CI}
% \textbf{TCP in CI.}
The existing work on TCP in CI can also be classified into techniques~\cite{liang2018redefining,haghighatkhah2018test,elbaum2014techniques,spieker2017reinforcement,bertolino2020learning,busjaeger2016learning} and experimental studies~\cite{hemmati2017prioritizing,haghighatkhah2018test,jin2021helped}. TCP techniques use heuristic strategies~\cite{elbaum2014techniques,haghighatkhah2018test,marijan2013test} or machine learning models~\cite{spieker2017reinforcement,busjaeger2016learning,sharif2021deeporder,bertolino2020learning,do2020multi} to predict the priority value of each test and schedule them accordingly. We will discuss these techniques in detail in Section~\ref{sec:TCP-select}. Besides these techniques, a series of empirical studies~\cite{hemmati2017prioritizing,haghighatkhah2018test,jin2021helped,yaraghi2022scalable,elsner2021empirically,bertolino2020learning} have been proposed. In particular, Hemmati et al.~\cite{hemmati2017prioritizing} conducted an experiment with three black-box TCP techniques on Firefox and found that historical information is essential for TCP in CI. Haghighatkhah et al.~\cite{haghighatkhah2018test} conducted an experiment to compare two categories of heuristic-based TCP, i.e., history-based TCP and diversity-based TCP in CI. Bertolino et al.~\cite{bertolino2020learning} conducted an experiment to compare learning-to-rank with ranking-to-learn strategies for TCP in CI. However, there are design flaws in their metric~\cite{pan2022test} and some recent ML-based techniques~\cite{bagherzadeh2021reinforcement,sharif2021deeporder,do2020multi} are not included.
Bagherzadeh et al.~\cite{bagherzadeh2021reinforcement} conducted an experiment on RL-based algorithms in CI, by combining them with three alternative ranking models, and regarded ACER-PA as the best prioritization technique. Jin and Servant~\cite{jin2021helped} compared 10 techniques using selection and prioritization strategies in CI and analyzed the benefit of different design decisions. 
They targeted a much broader scope, i.e., comparing selection and prioritization techniques at the build and test levels to find out what design decision brings benefits. However, our work focuses on a specialized scope, ML-based test prioritization, and aims to provide valuable insights for improving them through a fair comparison study. Jin and Servant’s work only included four prioritization techniques, which are all heuristic-based techniques. In contrast, our work investigates recent ML-based techniques, which are more advanced and perform better.
Elsner et al.~\cite{elsner2021empirically} evaluated SL-based and heuristic-based TCP techniques exclusively rely on metadata from VCS and CI systems as the metadata are readily available and inexpensive. Ling et al.~\cite{ling2021different} conducted an empirical study comparing nine heuristic-based and SL-based TCP techniques. Yaraghi et al.~\cite{yaraghi2022scalable} investigated the performance of one specific SL-based technique MART by considering the influence of data collection time, ML models, features, and the decay of effectiveness.
%Ling et al.~\cite{ling2021different} applied nine heuristic-based and SL-based TCP techniques to open and closed source projects and found that TCP techniques performed differently on different types of projects. Yaraghi et al.~\cite{yaraghi2022scalable} investigated the performance of SL-based technique by considering the influence of data collection time, ML-based models, features, and the decay of ML-based models. However, their conclusion is drawn based on the result of a single SL-based technique, MART.
% \dan{I modified this description. check and update if necessary.}
To sum up, the existing experiments have various goals, and the goal of this paper is different, i.e., a comprehensive study of ML-based TCP techniques in terms of effectiveness, efficiency, and applicability.

\section{Study Design}
% \vspace{-0.1cm}
In this section, we present the design details of this study. 
%We first discuss the selection of TCP techniques and the metrics used to measure their performance. Then we present the research questions we aim to answer. Finally, we present the basic information of the subjects and the implementation for the selected TCP techniques.
% \vspace{-0.2cm}
\subsection{TCP Technique Selection}
\label{sec:TCP-select}
% \vspace{-0.1cm}

To address the problem of TCP in CI, researchers have recently put dedicated efforts into this domain and have proposed many techniques~\cite{elbaum2014techniques,haghighatkhah2018test,marijan2013test,spieker2017reinforcement,busjaeger2016learning,sharif2021deeporder,bertolino2020learning,do2020multi}.
The first attempt starts with heuristics-based techniques, e.g., Elbaum et al.~\cite{elbaum2014techniques} prioritized tests based on whether they have not been executed for long or have failed in the recent commits, Haghighatkhah et al.~\cite{haghighatkhah2018test} scheduled tests based on the combination of their previous execution results and similarity. In this paper, we only consider ML-based techniques because they have better effectiveness and lower overhead (shown in recent research~\cite{spieker2017reinforcement,sharif2021deeporder,yaraghi2022scalable}). We select all the state-of-the-art ML-based TCP techniques for CI mainly by considering the review~\cite{pan2022test}. Moreover, some new approaches published after the review are also included.

\subsubsection{Inclusion}
We include 11 TCP techniques from previous work, which can be classified from two orthogonal aspects. First, they can be classified based on the learning strategy: SL-based techniques and RL-based techniques. 
% SL-based techniques first produce labels for each test case (e.g., 1 for a failing test and 0 for a passing test) and then train neural networks to predict the labels. 
SL-based techniques train neural networks to predict scores of tests (e.g., 1/0 for a failing/passing test).
% \dan{``produce''? misleading. since the technique has already produced the label, it is not necessary to train a model to predict the label.}
Then they treat the prediction results as priority values and produce prioritization results accordingly. Note that for supervised learning, the neural model gets the training data at once, using data batch to accelerate and stabilize training. Unlike SL-based techniques, RL-based techniques use predefined reward signals to guide the update of the prioritization strategy. The reward signal is calculated based on the execution result of a test or the distance between the predicted rank and the optimal rank of the test. 
% RL-based techniques reinforce the current strategy when getting positive rewards while exploring other strategies when getting negative rewards. 
% Besides,
% they continuously adjust the strategy during the interaction with the environment, i.e., 
RL-based techniques continuously update the strategy based on the coming data in each cycle. Second, the techniques can be classified based on the comparing strategy: pointwise, pairwise, and listwise ranking. The pointwise ranking strategy predicts scores for each single test. The strategy then gets the prioritization result by sorting the tests according to their predicted scores. The pairwise ranking strategy orders a pair of tests at a time. The strategy then gets the prioritization result using all the ordered pairs. The listwise ranking strategy orders a complete list of tests at a time, i.e., it takes in all the tests and directly produces the prioritization result. 
% Table \ref{tab:classify} lists the categorized techniques from the two aspects. 
Table \ref{tab:input} lists the selected techniques along with their features. 
We use ``SL'' and ``RL'' to represent SL-based and RL-based technique respectively. We use ``PO'', ``PA'', ``LI'' to represent pointwise, pairwise, listwise strategy respectively~\cite{bagherzadeh2021reinforcement}.
In the following, we present the details of the selected TCP techniques. 

\textbf{MART, RankNet, RankBoost, CA, L-MART:} Bertolino et al.~\cite{bertolino2020learning} study seven supervised learning algorithms (MART, RankNet, Rankboost, CA, L-MART, KNN, RF) and three reinforcement learning algorithms (RL, RL-MLP, RL-RF) on TCP in CI. Among them, MART, RankBoost, L-MART are ensemble algorithms and MART is shown to have the best performance. Besides MART, our study includes other competitive SL-based techniques, i.e., RankNet, Rankboost, CA, and L-MART. We exclude RL-MLP and RL-RF because they are reported~\cite{bagherzadeh2021reinforcement} to be less effective than the following RL-based techniques.
% \dan{why?}
% Instead, we include more advanced RL-based techniques as follows.

\textbf{ACER-PA, PPO2-PO, PPO1-LI:} Bagherzadeh et al.~\cite{bagherzadeh2021reinforcement} propose to apply ten state-of-the-art reinforcement learning (RL) algorithms (A2C, PPO1, PPO2, TRPO, DQN, ACKTR, ACER, DDPG, TD3, SAC) to TCP in CI. These RL algorithms are applied to three alternative ranking models, respectively: listwise ranking, pairwise ranking, and pointwise ranking.
% , resulting in 21 TCP techniques (lower than 30, because some algorithms cannot be applied to all the ranking models). 
In the paper~\cite{bagherzadeh2021reinforcement}, ACER-PA (i.e., ACER algorithm applied to pairwise ranking model) is the best and the recommended technique. In our study, besides ACER-PA, we also include the best pointwise technique (PPO2-PO) and the best listwise technique (PPO1-LI) to get a more comprehensive view.

\textbf{RETECS:} RETECS is the first RL-based technique~\cite{spieker2017reinforcement}, which builds a shallow neural network to prioritize tests based on the failure history, timestamp for previous execution, and approximated execution duration. We implement RETECS and use ``RL'' to refer to it following previous study~\cite{bertolino2020learning}.

\textbf{DeepOrder:} DeepOrder~\cite{sharif2021deeporder} is a SL-based technique. It trains a deep neural network to predict each test's priority value. The priority label is calculated by a heuristic technique, ROCKET~\cite{marijan2013test}. DeepOrder uses processed test history from previous CI cycles to better capture the feature of tests. 

\textbf{COLEMAN:} COLEMAN \cite{do2020multi} is a RL-based technique. It formulates tests as bandits and calculates the expected gain $Q$ for each bandit, which is then treated as the priority value in prioritization. The paper proposes two policies and two reward functions for COLEMAN. We use the best configuration for COLEMAN as shown in the paper, i.e., the combination of FRRMAB policy with TimeRank as the reward function.

% \begin{table}[!htp]
% \vspace{-0.4cm}
% \footnotesize
% \centering
% \caption{Categorized Techniques}
% \vspace{-0.2cm}
% \begin{tabular}{c|c|c}
% \hline
%   & \text{Supervised learning} & \text{Reinforcement learning} \\
% \hline \text { Pointwise } & DeepOrder & \makecell[c]{RETECS, COLEMAN,\\ PPO2-PO}\\
% \hline \text { Pairwise } & \makecell[c]{MART, RankNet, Rankboost,\\ L-MART} & ACER-PA\\
% \hline \text { Listwise } & CA & PPO1-LI\\
% \hline
% \end{tabular}
% \label{tab:classify}
% \vspace{-0.3cm}
% \end{table}

% \begin{table}[!htp]
% \vspace{-0.4cm}
% \footnotesize
% \centering
% \caption{The Selected Techniques}
% \vspace{-0.2cm}
% \begin{tabular}{c|c}
% \hline \text{Approach} & \text{Feature} \\
% \hline \text { \makecell[l]{MART, RankNet,\\RankBoost, CA, \\L-MART, RL, \\PPO2-PO, \\ACER-PA,\\ PPO1-LI }} & \makecell[l]{Number of test methods in a test class,\\
% Failure history of test cases,\\
% Previous execution time,\\
% Time span since the test case's last execution,\\
% Code characteristics}\\
% \hline \text { DeepOrder } & \makecell[l]{Failure history of test cases,\\
% Timestamp for the test case's last execution,\\
% Approximated execution time of test cases,\\
% Heuristically processed failure history}\\
% \hline \text { COLEMAN } & \makecell[l]{Number of test methods in a test class,\\
% Failure history of test cases,\\
% Timestamp for the test case's last execution,\\
% Approximated execution time of test cases}\\
% \hline
% \end{tabular}
% \label{tab:input}
% \vspace{-0.3cm}
% \end{table}

\begin{table}[!htp]
\vspace{-0.5cm}
\footnotesize
\centering
\caption{The Selected Techniques}
\vspace{-0.2cm}
\begin{tabular}{c|c}
\hline \text{Approach} & \text{Feature} \\
\hline \text { \makecell[l]{MART (SL, PA), \\RankNet (SL, PA),\\RankBoost (SL, PA), \\CA (SL, LI), \\L-MART (SL, PA), \\RL (RL, PO), \\PPO2-PO (RL, PO), \\ACER-PA (RL, PA),\\ PPO1-LI (RL, LI) }} & \makecell[l]{Number of test methods in a test class,\\
Failure history of test cases,\\
Previous execution time,\\
Time span since the test case's last execution,\\
Code characteristics}\\
\hline \text { DeepOrder (SL, PO) } & \makecell[l]{Failure history of test cases,\\
Timestamp for the test case's last execution,\\
Approximated execution time of test cases,\\
Heuristically processed failure history}\\
\hline \text { COLEMAN (RL, PO) } & \makecell[l]{Number of test methods in a test class,\\
Failure history of test cases,\\
Timestamp for the test case's last execution,\\
Approximated execution time of test cases}\\
\hline
\end{tabular}
\label{tab:input}
\vspace{-0.5cm}
\end{table}

% \vspace{-0.2cm}
\subsubsection{Exclusion} In our study, we discard some techniques. Their details and why they are excluded are given below.

Busjaeger et al. proposed a TCP technique based on machine learning\cite{busjaeger2016learning} and applied it in a large-scale CI environment. They extracted useful test features and used an SVM classifier to give priority to the tests. We discard this technique because it takes coverage information as input, which is expensive to get in the CI scenario. Besides, the authors do not publish their code and dataset. Yang et al. \cite{yang2020systematic} conducted a systematic study on the reward function of RL for TCP in CI. They explored the effects of the three reward functions, APHF, HFC, TF, and proposed to use time window in reward calculation to improve efficiency. 
% In addition, They also propose a new reward strategy, fuzzy reward strategy.
We discard this technique because its network structure is identical to that of RETECS, with the only variation being the utilization of different reward functions.
% We discard this technique because the network structure of their technique is the same as RETECS. The only difference is that they use different reward functions. 
Besides, the authors do not publish their code, either.

% \vspace{-0.2cm}
\subsection{Metrics Selection}
\label{sec:metric_selection}
% \vspace{-0.1cm}
Before exploring and comparing the performance of existing techniques, it is critical to decide how to measure the applicability and effectiveness of TCP in CI. In terms of applicability, we use Normalized Time Reduction (NTR)~\cite{do2020multi} to measure the ratio of the reduction time to the total execution time. In this metric, only failing CI cycles (i.e., CI cycles containing failing tests) are considered. It is computed as follows:

\setlength{\abovedisplayskip}{0.3\baselineskip}
\setlength{\belowdisplayskip}{0.3\baselineskip}
\begin{equation}
% \vspace{-0.1cm}
\small
\label{equ:ntr}
    \operatorname{NTR}(s)=\frac{\sum_{t=1}^{CI^{fail}}\left(\hat{r}_{t}-r_{t}\right)}{\sum_{t=1}^{CI^{fail}}\left(\hat{r}_{t}\right)}
% \vspace{-0.2cm}
\end{equation}

where $CI^{fail}$ represents the number of failing CI cycles, $\hat{r}_{t}$ represents the total execution time of all the tests in a cycle, $r_{t}$ represents the time spent until the first test fails. 

In terms of effectiveness, three metrics APFD, NAPFD, and NRPA have been used in existing work~\cite{bagherzadeh2021reinforcement,bertolino2020learning,spieker2017reinforcement}. Therefore, we will first introduce them in Section~\ref{sec:existing_metric}, then analyze their suitability in CI context in Section~\ref{sec:analysis_metric}.
% , and finally present a new metric in Section~\ref{sec:new_metric}.
% \vspace{-0.2cm}
\subsubsection{Existing Metrics}
\label{sec:existing_metric}

APFD measures the average percentage of faults detected by the scheduled tests~\cite{rothermel1999test} as follows:
\begin{equation}
% \vspace{-0.2cm}
\small
\label{equ:apfd}
    APFD\left(s\right)=1-\frac{\sum_{t \in s} rank\left(s, t\right) * t.v}{\left|s\right|* m}+\frac{1}{2*\left|s\right|}
% \vspace{-0.1cm}
\end{equation}
where $m$ represents the total number of faults, $t.v$ represents the verdict of test $t$ ($t.v=1/0$ means $t$ failed/passed), and $rank(s,t)$ represents the position of test $t$ in the scheduled test sequence $s$. A larger APFD value indicates that the failing tests are scheduled ahead. As a variant of APFD, NAPFD is proposed when some tests are not executed due to time constraints~\cite{qu2007combinatorial}. It normalizes the APFD value by including the percentage of faults detected by executed tests. As this paper does not investigate the performance of TCP in CI with time truncation, we do not consider NAPFD in this paper. 

The most used metric for TCP in CI is NRPA~\cite{bertolino2020learning,bagherzadeh2021reinforcement}, i.e., normalized rank percentile average.
Different from APFD and NAPFD, NRPA is designed to measure how close a scheduled test sequence $s$ is to the optimal sequence $s_o$.
% Given the optimal test sequence $s_o$, NRPA measures the scheduled test sequence $s$ through its normalized RPA values. 
The RPA value of a scheduled test sequence $s$ is computed as follows:
\begin{equation}
\label{equ:rpa}
\small
% \vspace{-0.2cm}
    RPA(s)=\frac{\sum_{t \in s} \sum_{i=rank(s, t)}^{k}|s|-rank\left(s_{o}, t\right)+1}{k^{2}(k+1)/2}
% \vspace{-0.1cm}
\end{equation}
where $rank(s,t)$ denotes the position of test $t$ in $s$, $k$ denotes the total number of tests in the sequence. NRPA is obtained by normalizing the RPA value, i.e., $NRPA(s)=\frac{RPA(s)}{RPA(s_o)}$, which ranges from 0 to 1. 
% Different from APFD and NAPFD, a smaller NRPA value indicates the produced test sequence is closer to the optimal test sequence.
A larger NRPA value indicates that the scheduled test sequence is closer to the optimal one. Note that APFD and NAPFD can only be computed in failing CI cycles, but NRPA can be computed in all CI cycles. For NRPA, in CI cycles with no failing tests, the optimal sequence means the passing tests are ranked based on the ascending order of their execution time.

\subsubsection{Analysis on Metrics}
\label{sec:analysis_metric}

Although being widely used, both NRPA and APFD have design flaws in measuring TCP in CI.

NRPA treats failing and passing tests as equally important and has unsatisfactory discernment. First, NRPA only focuses on the rank of tests, ignoring their execution results. Given an optimal sequence [5,4,3,2,1] (each number denotes the test's priority value) and a sequence scheduled by a TCP technique [5,2,3,4,1], NRPA only cares about the index, ignoring the tests' verdicts, i.e., we get NRPA=0.93 for the scheduled sequence regardless of execution results. However, if the tests with the priority of 2 and 4 are both passing, the scheduled sequence has the same fault detection ability as the optimal sequence. That is, the metric fails to evaluate sequences based on their ability to detect faults. Second, NRPA has unsatisfactory discernment. If the tests with the priority of 2 and 4 are passing and failing respectively, the scheduled sequence [5,2,3,4,1] and the optimal sequence [5,4,3,2,1] have a large difference in fault detection ability, but the NRPA value of the former sequence (i.e., 0.93) is close to 1, indicating that it is close to the optimal. Moreover, NRPA can be misleading in certain scenarios~\cite{pan2022test}, with a sequence having a higher NRPA value potentially being worse. Therefore, we do not adopt NRPA as the measurement.

Compared with NRPA, APFD is more reliable because it only considers the ranks of failing tests. However, it cannot power fair comparison across CI cycles because it has different value ranges across CI cycles. Let us suppose each failing test accounts for a distinct fault. In the best situation where all failing tests are prioritized ahead, we get $APFD_{max}(s)=1-\frac{1+2+\cdots+m}{|s|*m}+\frac{1}{2|s|}=1-\frac{m}{2|s|}$, while in the worst situation, we get $APFD_{min}(s)=1-\frac{(|s|-m+1)+\cdots+|s|}{|s|*m}+\frac{1}{2|s|}=\frac{m}{2|s|}$. Therefore, APFD may have different upper bound and lower bound in different CI cycles, depending on the number of tests ($|s|$) and faults ($m$) in the test suite. Therefore, APFD values in different cycles can hardly be compared.
% Therefore, APFD values in each cycle can hardly be compared and It is difficult to judge whether the TCP technique performs better or worse in each cycle. 
% APFD cannot reflect the change of the technique's effectiveness across CI cycles. 
Moreover, APFD values cannot reflect the distance between the scheduled and optimal sequences. To address these issues, we normalize the AFPD value with min-max normalization, resulting in the following rectified APFD (rAPFD), used as our metric. 
\begin{equation}
\small
% \vspace{-0.1cm}
    rAPFD(s) = \frac{APFD(s)-APFD_{min}(s)}{APFD_{max}(s)-APFD_{min}(s)}
% \vspace{-0.1cm}
\end{equation}
For various cycles, rAPFD has the same value range $[0,1]$. $rAPFD=1$ indicates that the scheduled sequence is optimal.

\subsection{Research Questions}
% \vspace{-0.2cm}
\label{sec:rq}

This paper studies ML-based TCP techniques in CI from three aspects: effectiveness, efficiency, and applicability. %To achieve this goal, we aim to answer the following research questions.

\textbf{RQ1. Effectiveness comparison.} The existing techniques have not been fully evaluated under the same experimental setup. Therefore, we compare their effectiveness and propose two minor research questions.

% \textbf{RQ1.1. How do the ML-based techniques perform across subjects?} This research question aims to examine whether ML-based TCP techniques exhibit different performance on different subjects and to identify the reasons for any observed difference. To answer this question, we compute the rAPFD results of the studied TCP techniques on each subject. If we observe that the techniques perform differently on different subjects, we will attempt to explain the difference and conduct additional experiments to validate the explanation. Answering this question can help us choose an appropriate TCP technique when facing different kinds of subjects. 

\textbf{RQ1.1. How do the ML-based techniques perform across subjects?} This research question aims to examine whether ML-based TCP techniques exhibit different performance on different subjects and to identify the reasons for any observed difference. To answer this question, we compute the rAPFD results of the studied TCP techniques on each subject. Answering this question can help us choose an appropriate TCP technique when facing different kinds of subjects. 

% \textbf{RQ1.2. How do the ML-based techniques perform across CI cycles?} This research question aims to examine whether ML-based TCP techniques exhibit different performance across different CI cycles and to identify the reasons for any observed difference. 
% To answer this question, we plot the trend of the rAPFD results on the subjects based on the timeline. If we observe that the TCP techniques perform differently on earlier and later cycles, we will attempt to explain the difference and conduct additional experiments to prove the explanation. Answering this question can help us understand the effect of continuously coming CI data on the effectiveness of TCP techniques.

\textbf{RQ1.2. How do the ML-based techniques perform across CI cycles?} This research question aims to examine whether ML-based TCP techniques exhibit different performance across CI cycles and the reasons behind it. To answer this question, we plot the trend of the rAPFD results on the subjects based on the timeline. Answering this question can help us understand the effect of continuously coming CI data on the effectiveness of TCP techniques.

\textbf{RQ2. Efficiency Comparison.}
Time cost plays an important role in choosing techniques because TCP is proposed to speed up fault detection during testing. This research question aims to study the efficiency of ML-based TCP techniques, in terms of the time spent for model learning (i.e., training time) and the time spent for scheduling tests (i.e., prediction time). 

\textbf{RQ3. Applicability.}
In this RQ, we aim to investigate if the techniques are applicable in practice. If the training or prediction time is longer than the commit interval, the technique cannot complete training or produce a prioritization result before the next commit comes, which makes it inapplicable. Therefore, we compare the average commit interval of each subject with the techniques' training and prediction time. Besides, a large overhead makes TCP unnecessary if simply executing all the tests costs less time. Therefore, we compare the total execution time of the tests with the prediction time of the techniques. What is more, we would like to know with TCP techniques how much time could be saved. Therefore, we compute the NTR results of the studied TCP techniques.

\begin{table*}[tp]
\centering
\small
\setlength{\abovecaptionskip}{0cm} 
\setlength{\belowcaptionskip}{-0.2cm}
\caption{Statistics for subjects}
\resizebox{\linewidth}{!}{
% \begin{tabular}{c|ccccccccccc}
% \hline \text {Subjects} & \text{bcel} & \text{csv} & \text{dbcp} & \text{text} & \text{java-faker} & \text{jedis} & \text{jsoup} & \text{jsprit} & \text{maxwell} & \text{nfe} & \text{spring-data-redis} \\
% \hline \text { SLOC } & 165,246 & 168,508 & 57,091 & 52,033 & 66,313 & 60,782 & 40,206 & 192,861 & 36,907 & 62,826 & 227,697 \\
% \hline \text { Selected classes } & 2,860 & 2,913 & 5,159 & 2,872 & 11,645 & 15,019 & 7,346 & 3,748 & 4,687 & 19,833 & 7,216\\
% \hline \text { Failing classes } & 79 & 5 & 19 & 5 & 15 & 290 & 11 & 177 & 8 & 1,996 & 476\\
% \hline \text { CI cycle } & 177 & 270 & 324 & 224 & 217 & 446 & 378 & 75 & 231 & 185 & 177\\
% \hline \text { Failing cycle } & 79 & 5 & 5 & 3 & 14 & 214 & 8 & 27 & 6 & 106 & 114\\
% \hline \text { Failure rate (\%) } & 2.7622 & 0.1716 & 0.3683 & 0.1741 & 0.1288 & 1.9309 & 0.1497 & 4.7230 & 0.1707 & 10.0640 & 6.5965\\
% \hline \text { Avg commit interval (s) } & 12,756.52 & 15,660.48 & 11,236.82 & 16,955.01 & 20,099.52 & 25,343.38 & 11,514.23 & 17,370.03 & 20,943.48 & 18,304.21 & 18,490.62\\
% \hline \text { Test duration (s) } & 0.065 $\sim$ 28.376 & 6.639 $\sim$ 62.261 & 11.627$\sim$124.718 & 0.212$\sim$272.123 & 0.104 $\sim$ 131.940 & 0.170 $\sim$ 6.622 & 0.102 $\sim$ 52.939 & 1.315 $\sim$ 32.412 & 3.273 $\sim$ 1,228.466 & 0.13 $\sim$ 2,123.381 & 1.298 $\sim$ 431.736\\
% \hline
% \end{tabular}
\begin{tabular}{c|ccccccccccc}
\hline 
\text{Subjects} & \text{SLOC} & \text{Selected classes} & \text{Failing classes} & \text{CI cycle} & \text{Failing CI cycle} & \text{Failure rate (\%)} & \text{Avg commit interval (s)} & \text{Test duration (s)} & Time period (months)\\
\hline
\text{bcel} & 165,246 & 2,860 & 79 & 177 & 79 & 2.7622 & 12,756.52 & 0.065 $\sim$ 28.376 & 69\\
\text{csv} & 168,508 & 2,913 & 5 & 270 & 5 & 0.1716 & 15,660.48 & 6.639 $\sim$ 62.261 & 86\\
\text{dbcp} & 57,091 & 5,159 & 19 & 324 & 5 & 0.3683 & 11,236.82 & 11.627$\sim$124.718 & 62\\
\text{text} & 52,033 & 2,872 & 5 & 224 & 3 & 0.1741 & 16,955.01 & 0.212$\sim$272.123 & 34\\
\text{java-faker} & 66,313 & 11,645 & 15 & 217 & 14 & 0.1288 & 20,099.52 & 0.104 $\sim$ 131.940 & 87 \\
\text{jedis} & 60,782 & 15,019 & 290 & 446 & 214 & 1.9309 & 25,343.38 & 0.170 $\sim$ 6.622 & 78\\
\text{jsoup} & 40,206 & 7,346 & 11 & 378 & 8 & 0.1497 & 11,514.23 & 0.102 $\sim$ 52.939 & 121\\
\text{jsprit} & 192,861 & 3,748 & 177 & 75 & 27 & 4.7230 & 17,370.03 & 1.315 $\sim$ 32.412 & 83\\
\text{maxwell} & 36,907 & 4,687 & 8 & 231 & 6 & 0.1707 & 20,943.48 & 3.273 $\sim$ 1,228.466 & 32\\
\text{nfe} & 62,826 & 19,833 & 1,996 & 185 & 106 & 10.0640 & 18,304.21 & 0.13 $\sim$ 2,123.381 & 37\\
\text{spring-data-redis} & 227,697 & 7,216 & 476 & 177 & 114 & 6.5965 & 18,490.62 & 1.298 $\sim$ 431.736 & 44 \\
\hline
\end{tabular}
}
\vspace{-0.4cm}
% \label{table-collected-dataset1}
\label{tab:sub}
\end{table*}

% \vspace{-0.2cm}
\subsection{Subjects}
% \vspace{-0.1cm}
\label{sec:rq}

In this study we construct a new dataset based on 11 highly-starred, open-source subjects from GitHub, since the existing datasets~\cite{bagherzadeh2021reinforcement,bertolino2020learning,yaraghi2022scalable,elsner2021empirically} do not contain all features of the studied techniques (given by Table~\ref{tab:input}). We select the top 11 highly-starred projects with restrictions on program size ($>10k$ SLOC), the number of commits ($>800$), and the number of test classes per commit ($>10$). In particular, four of them (i.e., bcel, csv, dbcp, text) are Apache Commons projects and the rest are popular large-scale projects. %They all use Maven as their build system since Maven provides a standard build and testing framework with detailed output traces. 
The four Apache projects have been deployed on Github CI while the other projects have been deployed on Travis CI, which is exactly the application scenario we target at.

As existing work~\cite{bertolino2020learning} does, we collect the CI data as follows. For each subject, we collect the latest 800 commits, including the source code and the test suite for each commit, regarded as 800 CI cycles. Therefore, in total we obtain 8,800 versions of the 11 subjects, each with a corresponding test suite. In each CI cycle, we extract files affected by the changed code and select the test classes covering the changed files. In other words, we remove the CI cycles without selected test classes and prioritize only selected test classes. 
Table~\ref{tab:sub} presents the statistics of the 11 subjects, where Column ``Failure rate'' presents the ratio of failing test classes. Column ``Avg commit interval'' presents the average commit interval for the 800 commits. Column ``Test duration'' presents the range of the regression testing time in each cycle. Note that the duration varies a lot in a single subject, because we only consider the selected test classes in each cycle.

% \vspace{-0.2cm}
\subsection{Implementation}
We implement all the techniques using their default parameters and strictly use the features proposed in their papers~\cite{bagherzadeh2021reinforcement,bertolino2020learning,sharif2021deeporder,do2020multi} as input.
%, i.e., different techniques use different features as they were originally proposed and tuned with varying feature sets. The inputs are part of their model design, which affects their performance and application scenario.} 
More specifically, we use Ranklib library~\cite{Ranklib} to implement MART, L-MART, RankNet, CA, and RankBoost~\cite{bertolino2020learning}, while using the published code to implement the other techniques. Besides, we use Understand~\cite{SciToolsUnderstand} to obtain the static code features of the studied TCP techniques. When calculating rAPFD, we regard each failing test as revealing a unique fault.

For each subject, we use the first 2000 test classes of the selected 800 cycles as training instances
% \footnote{If tests within one cycle are divided into training and testing sets, in such case we put all the tests within the cycle into the training set. In other words, sometimes a subject may have more than 2000 training instances.} 
and the remaining as testing instances following previous work~\cite{bertolino2020learning}. In other words, we select as many CI cycles until reaching 2000 test classes and treat them as training cycles. Regarding the testing instances, in each cycle, each compared ML-based TCP technique produces a scheduled sequence of test classes. We then evaluate the sequence using the selected metric. All the experiments are conducted on a server with Ubuntu 16.04 x64 OS, 2 Intel Xeon E5-2680 v4 CPUs, and 377GB memory.  

% \vspace{-0.2cm}
\section{Results and Analysis}

In this section, we present and analyze the results.

\begin{figure}
  \centering
  \includegraphics[width=0.4\textwidth]{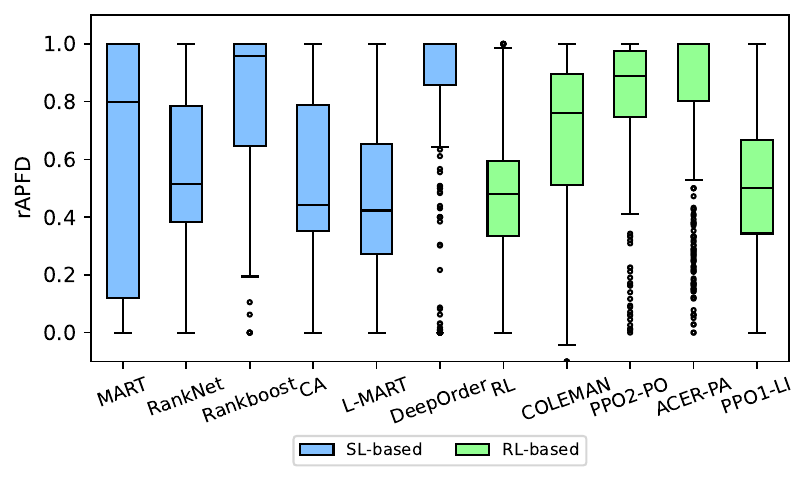} 
  % \vspace{-0.4cm}
  \caption{rAPFD for each TCP technique across all failing cycles}
  \label{fig-total} 
  \vspace{-0.4cm}
\end{figure}

\begin{table*}[!htp]
\centering
\footnotesize
\begin{threeparttable}
\setlength{\abovecaptionskip}{0cm} 
\setlength{\belowcaptionskip}{-0.3cm}
\caption{Average rAPFD for compared TCP techniques}
% \vspace{-0.3cm}
\label{table:DDP}
\begin{tabular}{r|ccccc|cccccc}
\hline \multirow{2}{*}{Subjects} & \multicolumn{5}{c|}{More-failure subjects} & \multicolumn{6}{c}{Less-failure subjects} \\
\cline{2-12}  & \text{bcel} & \text{jedis} & \text{jsprit} & \text{nfe} & \text{spring-data-redis} &\text{csv} &\text{dbcp} & \text{text} & \text{java-faker} & \text{jsoup} & \text{maxwell} \\
\hline \text { MART (SL,PA) } & 0.7023 & 0.1930 & \textbf{1.0000} & \textbf{1.0000} & \textbf{0.9733} & \textbf{1.0000} & \textbf{1.0000} & \textbf{1.0000} & \textbf{0.9512} & \textbf{1.0000} & 0.6231 \\
\hline \text { RankNet (SL,PA) } & 0.4525 & \textbf{0.6835} & \textbf{0.7967} & 0.4220 & 0.5453 & \textbf{1.0000} & 0.3586 & \textbf{0.7143} & 0.2543 & 0.3095 & 0.1468 \\
\hline \text { RankBoost (SL,PA) } & \textbf{1.0000} & 0.6512 & \textbf{1.0000} & \textbf{0.9892} & \textbf{0.9908} & \textbf{0.6667} & 0.0839 & 0.2857 & \textbf{0.9589} & \textbf{0.8980} & \textbf{0.6960} \\
\hline \text { CA (SL,LI) } & 0.8205 & 0.3377 & 0.6693 & 0.5005 & \textbf{0.9582} & 0.6111 & 0.4161 & 0.4286 & 0.4881 & \textbf{0.7998} & \textbf{0.6837} \\
\hline \text { L-MART (SL,PA) } & 0.6255 & 0.3460 & 0.4769 & \textbf{0.8659} & 0.4044 & 0.2778 & 0.0526 & 0.4286 & 0.3385 & 0.4303 & 0.5938 \\
\hline \text { DeepOrder (SL,PO) } & \textbf{0.9680} & \textbf{0.9991} & 0.6547 & \textbf{0.9752} & 0.2120 & 0.2222 & \textbf{0.6447} & \textbf{0.7143} & 0.4712 & \textbf{0.6832} & \textbf{0.9460} \\
\hline \text { RL (RL,PO) } & 0.4792 & 0.4337 & 0.5171 & 0.5207 & 0.4875 & 0.5278 & \textbf{0.6694} & 0.4286 & \textbf{0.6455} & 0.3160 & \textbf{0.6828} \\
% \hline \text { RL-MLP } & 0.4578 & 0.5468 & 0.4943 & 0.4797 & 0.5015 & 0.1795 & 0.3997 & 0.2500 & 0.4877 & 0.2105 & 0.7626\\
\hline \text { COLEMAN (RL,PO) } & \textbf{0.8768} & \textbf{0.8031} & 0.0768 & 0.7034 & 0.3858 & \textbf{0.9722} & \textbf{0.8092} & 0.1429 & \textbf{0.5497} & \textbf{0.7416} & 0.3807\\
\hline \text { PPO2-PO (RL,PO) } & \textbf{0.9491} & \textbf{0.8770} & \textbf{0.7818} & 0.7249 & \textbf{0.8917} & 0.5556 & 0.2500 & \textbf{0.7143} & 0.2297 & 0.4840 & 0.0540 \\
\hline \text { ACER-PA (RL,PA) } & \textbf{1.0000} & \textbf{0.7963} & \textbf{0.7985} & \textbf{0.9471} & \textbf{0.9839} & 0.2500 & 0.3553 & 0.2857 & 0.3543 & 0.3540 & 0.0312 \\
\hline \text { PPO1-LI (RL,LI) } & 0.3872 & 0.4788 & 0.7705 & 0.5308 & 0.5024 & \textbf{0.7500} & \textbf{0.6431} & \textbf{1.0000} & \textbf{0.6450} & 0.4398 & \textbf{0.9545} \\
\hline
\end{tabular}

\end{threeparttable}
\vspace{-0.4cm}
\end{table*}

% \vspace{-0.2cm}
\subsection{RQ1: Effectiveness Comparison}
\label{sec:rq1}

\subsubsection{RQ1.1 Effectiveness across subjects}
\label{sec:rq1.1}
Fig.~\ref{fig-total} shows the rAPFD results of the TCP techniques across all failing cycles in all subjects, where blue boxes represent SL-based techniques and green boxes represent RL-based techniques. Among the five RL-based techniques, the boxes of ACER-PA and PPO2-PO are the highest, indicating that they are the most effective RL-based techniques.
Among the six SL-based techniques, the boxes of DeepOrder and RankBoost are the highest, indicating that they are the most effective SL-based techniques.
Furthermore, among all the compared 11 TCP techniques, the box of DeepOrder is the highest, indicating that it achieves the best effectiveness across all CI cycles. In terms of result distribution, MART exhibits the highest variance in rAPFD values, suggesting its potential instability across different CI cycles. Several approaches, such as DeepOrder, PPO2-PO, and ACRE-PA have many outliers, which also show their unstable performance.

\begin{figure}
  \centering
  \includegraphics[width=0.4\textwidth]{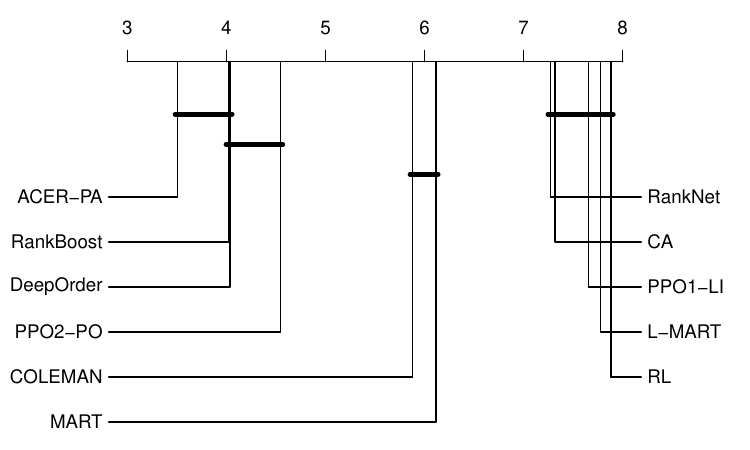} 
  % \vspace{-0.3cm}
  \caption{Critical difference plot}
  \vspace{-0.5cm}
  \label{fig-postHocTest} 
\end{figure}

To investigate whether these techniques perform significantly differently, we conduct a non-parametric ANOVA analysis utilizing the Friedman test with the Iman and Davenport extension~\cite{iman1980approximations}, which is robust to non-normality and heteroscedasticity. The analysis result rejects the null hypothesis of equal performance (with $p<2.2\times10^{-16}$), indicating that at least one TCP technique performs significantly differently from the others. Furthermore, to investigate which techniques perform significantly better, we run a post hoc test (i.e., Friedman post hoc test corrected by Shaffer's static procedure~\cite{calvo2016scmamp}) to compare each pair of TCP techniques. The analysis results are given by Fig.~\ref{fig-postHocTest}, where each technique is placed on an axis according to its mean rank among the 11 techniques across all outcomes. Therefore, TCP techniques with larger rAPFD are placed on the left. The TCP techniques not grouped with a horizontal line are significantly different ($p < 0.05$). A larger distance between two TCP techniques indicates their larger critical difference. From the figure, ACER-PA, RankBoost, DeepOrder, and PPO2-PO are in the best group, i.e., each of which significantly outperforms the techniques in the other groups.
% \dan{the techniques in the other group?}. 
The worst group includes RankNet, CA, PPO1-LI, L-MART, and RL. 
Note that in the best group, all techniques are pointwise and pairwise techniques, while in the worst group, all listwise techniques are included (i.e., CA and PPO1-LI). Therefore, we find that listwise techniques have relatively low effectiveness for TCP in CI, which is consistent with previous work~\cite{bagherzadeh2021reinforcement}. Listwise ranking is more challenging because it has a higher dimension of input, which grows linearly with the number of tests in a test suite. 
% \dan{I cannot understand ``difficult'' and ``large observation space''.}

% \vspace{-0.3cm}
% \subsubsection{Detailed analysis on each subject}
% \textbf{Detailed analysis on each subject}
We further present the average rAPFD results in Table~\ref{table:DDP}, where the 2nd to 6th columns present the subjects with a non-trivial number of failures (denoted as more-failure subjects, i.e., whose failure rate is larger than 1\%), and the 8th to 13th columns present the subjects with a trivial number of failures (denoted as less-failure subjects, i.e., whose failure rate is not larger than 1\%). We highlight the top-5 rAPFD results for each subject in bold font.
% and calculate their number of occurrences for each technique (given by Column ``Top-5''). 
In the following, we analyze the rAPFD results of the compared TCP techniques on more-failure subjects and less-failure subjects, respectively. On the more-failure subjects (i.e., \textit{bcel}, \textit{jedis}, \textit{jsprit}, \textit{nfe}, and \textit{spring-data-redis}), ACER-PA outperforms other RL-based techniques (on all the more-failure subjects, it is among the top-5 most effective techniques). RankBoost outperforms other SL-based techniques (on 4 out of 5 subjects, it is among the top-5 most effective techniques). Moreover, ACER-PA performs better than RankBoost, and thus it achieves the best performance on more-failure subjects. On the less-failure subjects (i.e., \textit{csv}, \textit{dbcp}, \textit{text}, \textit{java-faker}, \textit{jsoup}, and \textit{maxwell}), MART outperforms other SL-based techniques (on 5 out of 6 subjects, it is among the top-5 most effective techniques), while PPO1-LI outperforms other RL-based techniques (on 5 out of 6 projects, it is among the top-5 most effective techniques). Moreover, MART performs much better than PPO1-LI (on 4 out of 6 subjects, it produces the optimal sequence in each cycle, i.e., it has the rAPFD values of ``1'' on the 4 subjects). Therefore, MART achieves the best performance on less-failure subjects. For all subjects, MART is among the top-5 most effective techniques on 8 out of 11 subjects, therefore, MART has the best overall performance. Note that in Fig.~\ref{fig-total}, we analyze from the perspective of all failing cycles. Therefore, DeepOrder appears to have the best result because it performs the best on \textit{jedis}, which has the most failing cycles. However, from the perspective of subjects, we reach a different conclusion. 

\textbf{Besides, it is also worth noting that the best TCP technique on more-failure subjects, ACER-PA, performs the worst on less-failure subjects. The result is contradictory to previous work~\cite{bagherzadeh2021reinforcement}, where ACER-PA yields better results than MART on all subjects.} Here we analyze the contradictory conclusion between our work and the previous work~\cite{bagherzadeh2021reinforcement}. First, in the evaluation comparison conducted by the previous work, only less-failure subjects are used, each of which contains a small number of failing cycles ($<10$). Second, the previous work uses the combination of NRPA and APFD as the metric, i.e., NRPA results are reported for passing cycles (cycles without failing tests) while APFD results are reported for failing cycles. When it comes to the overall average performance, NRPA yields misleading values as explained in Section~\ref{sec:analysis_metric} and affects the average result to a large extent due to the large proportion of passing cycles.

\begin{figure}[tp]
    \centering
    % \text{\footnotesize SU}
    % \vspace{-0.4cm}
    % \setlength{\abovecaptionskip}{0.1cm} 
    % \setlength{\belowcaptionskip}{-0.1cm}
    % \subfigure{
    \subfigure{
        \includegraphics[width=0.23\textwidth]{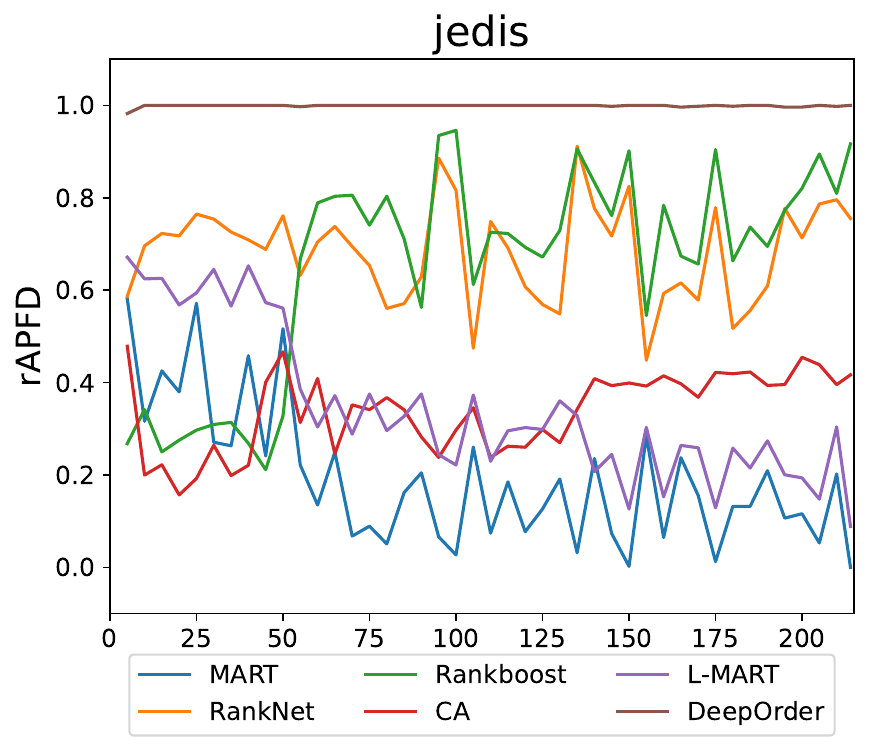}
    }
    \hspace{-3mm}
    \vspace{-3mm}
    \subfigure{
        \includegraphics[width=0.23\textwidth]{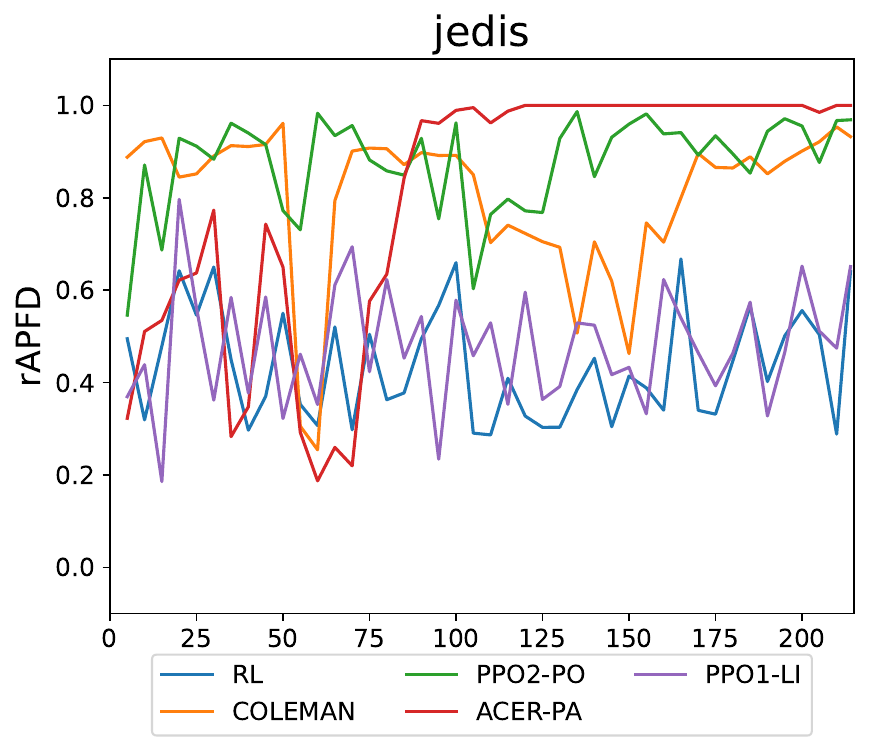}
    }
    \vspace{-3mm}
    \subfigure{
        \includegraphics[width=0.23\textwidth]{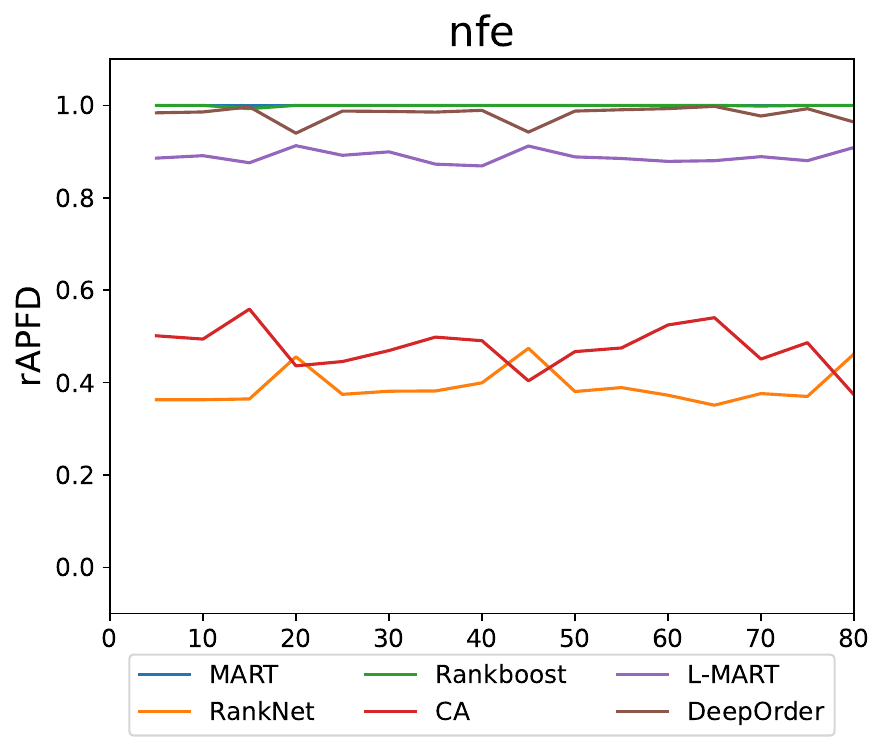}
    }
    \hspace{-3mm}
    \subfigure{
        \includegraphics[width=0.23\textwidth]{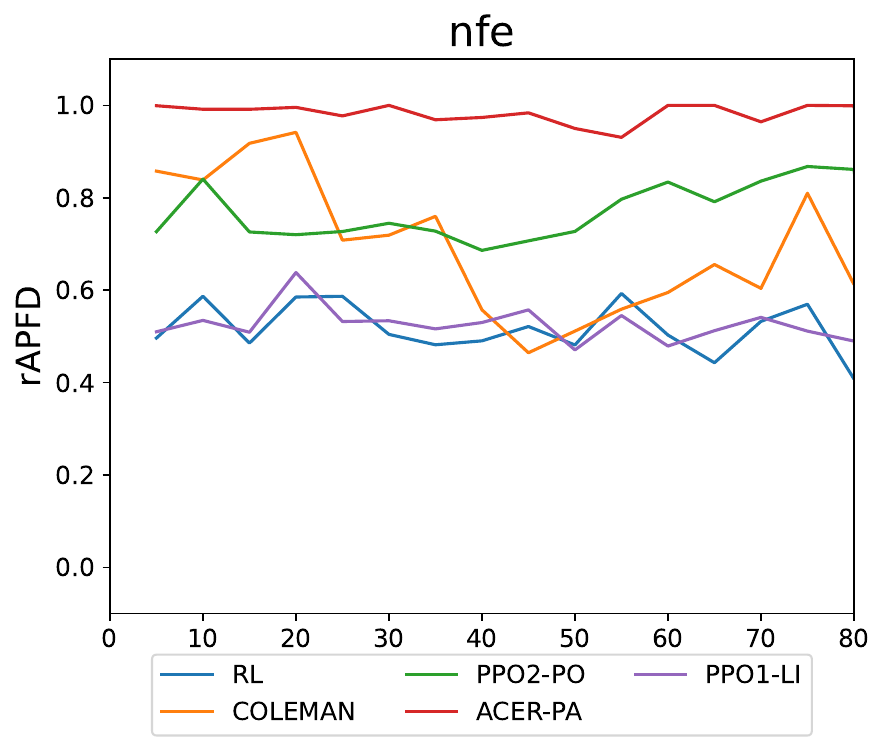}
    }
    \vspace{-3mm}
    \subfigure{
        \includegraphics[width=0.23\textwidth]{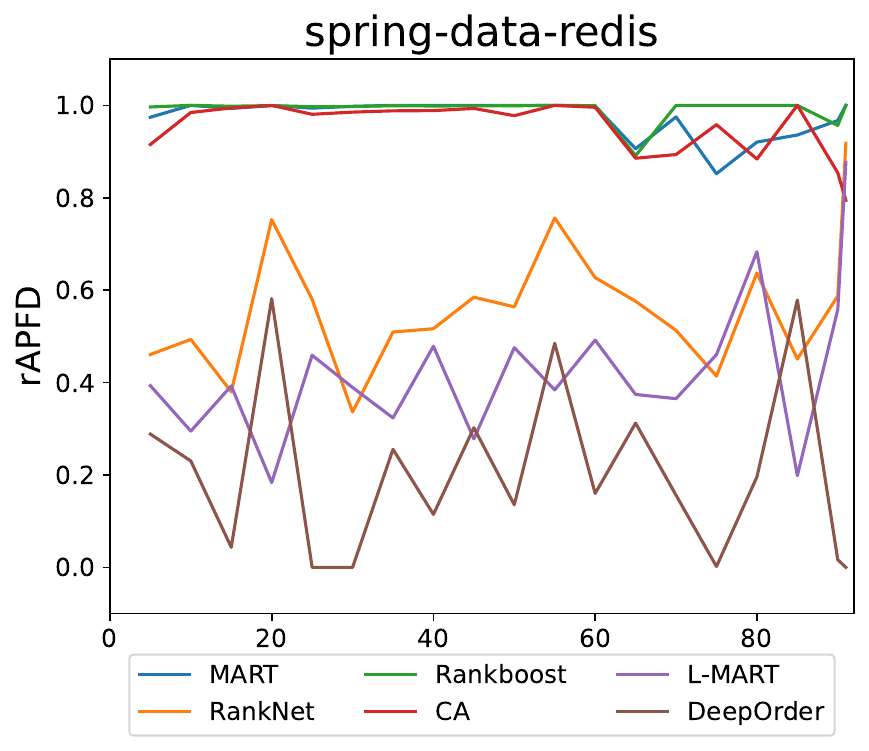}
    }
     \hspace{-3mm}
    \subfigure{
        \includegraphics[width=0.23\textwidth]{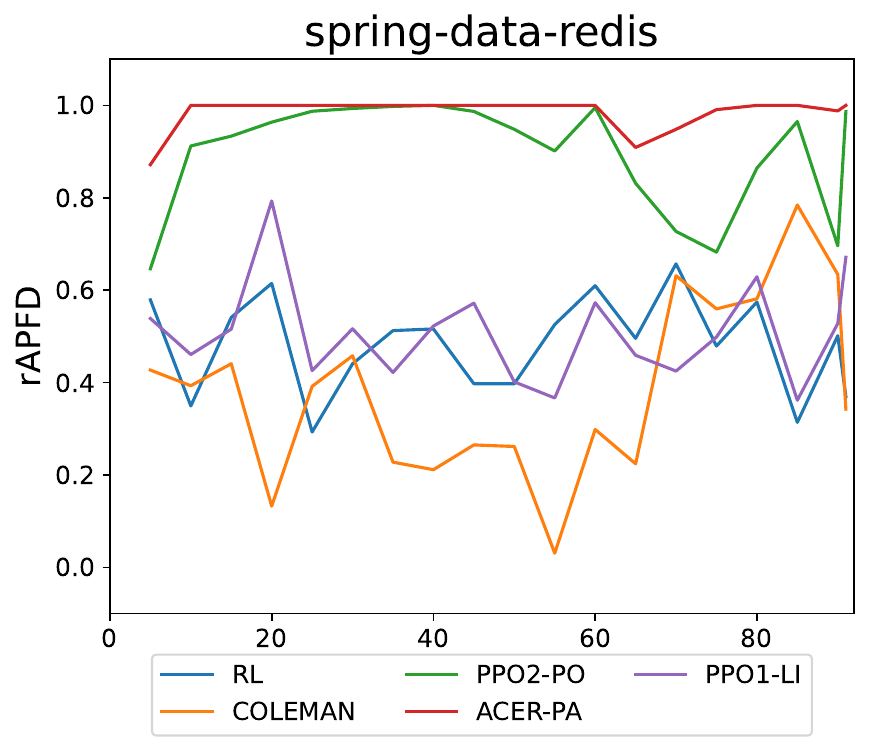}
    }
        
    \caption{Effectiveness change across CI cycles (the left column for SL-based techniques, the right column for RL-based techniques)}
    \label{fig-ci-cycle}
    \vspace{-0.3cm}
\end{figure}

% \vspace{-0.05cm}
From the preceding results, we get the following observations. First, within the two categories (i.e., SL-based and RL-based) of TCP techniques, techniques have different performance on more-failure and less-failure subjects. In the same category, the best technique is not the same for the two kinds of subjects. \textbf{Second, across the two categories of TCP techniques, RL-based technique ACER-PA performs the best on more-failure subjects, while SL-based technique MART performs the best on less-failure subjects. Overall, MART achieves the best effectiveness on all subjects.} 
Why do TCP techniques perform differently on more-failure and less-failure subjects?
% Why do TCP techniques have different performance on different kinds of subjects?

\begin{table*}[!htp]
\centering
\footnotesize
\begin{threeparttable}
\setlength{\abovecaptionskip}{0cm} 
\caption{Average rAPFD for compared TCP techniques after data pre-processing}
\label{table:DDP-smote}
\begin{tabular}{r|ccccc|cccccc}
\hline \multirow{2}{*}{Subjects} & \multicolumn{5}{c|}{More-failure subjects} & \multicolumn{6}{c}{Less-failure subjects} \\
\cline{2-12}  & \text{bcel} & \text{jedis} & \text{jsprit} & \text{nfe} & \text{spring-data-redis} &\text{csv} &\text{dbcp} & \text{text} & \text{java-faker} & \text{jsoup} & \text{maxwell}\\
\hline \text { MART (SL,PA) } & \textbf{0.9295} & \textbf{1.0000} & \textbf{1.0000} & \textbf{1.0000} & \textbf{0.9881} & \textbf{1.0000} & \textbf{1.0000} & \textbf{1.0000} & 0.7183 & \textbf{1.0000} & 0.2509 \\
\hline \text { RankNet (SL,PA) } & \textbf{0.5958} & \textbf{0.8702} & \textbf{1.0000} & \textbf{0.4353} & \textbf{0.7117} & \textbf{1.0000} & 0.3141 & \textbf{1.0000} & \textbf{0.7076} & 0.1952 & \textbf{0.6563} \\
\hline \text { RankBoost (SL,PA) } & \textbf{1.0000} & \textbf{0.8769} & \textbf{1.0000} & \textbf{0.9929} & 0.9900 & \textbf{1.0000} & \textbf{0.3141} & \textbf{0.5714} & \textbf{1.0000} & \textbf{1.0000} & \textbf{0.9773}\\
\hline \text { CA (SL,LI) } & 0.4341 & \textbf{0.9703} & \textbf{0.7428} & 0.0680 & 0.4000 & 0.0833 & \textbf{0.9737} & 0.0000 & \textbf{0.6354} & 0.5612 & 0.2188\\
\hline \text { L-MART (SL,PA) } & \textbf{1.0000} & \textbf{0.9659} & \textbf{0.5105} & \textbf{0.9131} & \textbf{0.9903} & \textbf{0.4444} & \textbf{1.0000} & 0.2857 & 0.1993 & \textbf{0.6433} & \textbf{1.0000}\\
\hline \text { DeepOrder (SL,PO) } & \textbf{1.0000} & 0.9625 & 0.1451 & 0.1593 & 0.1885 & 0.2222 & 0.6447 & 0.7143 & 0.4081 & 0.6832 & 0.9460\\
\hline \text { RL (RL,PO) } & \textbf{0.5172} & 0.4196 & 0.4137 & \textbf{0.5221} & \textbf{0.4975} & 0.3899 & \textbf{0.7253} & \textbf{0.5714} & 0.5083 & \textbf{0.4326} & 0.6591\\
% \hline \text { RL-MLP } & 0.4578 & 0.5468 & 0.4943 & 0.4797 & 0.5015 & 0.1795 & 0.3997 & 0.2500 & 0.4877 & 0.2105 & 0.7626\\
\hline \text { COLEMAN (RL,PO) } & 0.8505 & 0.7960 & \textbf{0.0870} & \textbf{0.8035} & 0.3546 & 0.5833 & 0.8092 & \textbf{0.8571} & 0.3679 & 0.5670 & \textbf{0.4839}\\
\hline \text { PPO2-PO (RL,PO) } & 0.8418 & 0.7536 & \textbf{0.8637} & \textbf{0.7651} & 0.8711 & \textbf{0.6944} & \textbf{0.3766} & 0.2857 & 0.2142 & 0.4093 & 0.0417\\
\hline \text { ACER-PA (RL,PA) } & 0.8889 & \textbf{0.8220} & \textbf{1.0000} & \textbf{0.9712} & \textbf{0.9912} & \textbf{0.4167} & \textbf{0.3816} & 0.2857 & \textbf{0.8922} & \textbf{0.5241} & \textbf{1.0000}\\
\hline \text { PPO1-LI (RL,LI) } & \textbf{0.5446} & 0.4693 & 0.6807 & \textbf{0.5354} & 0.4845 & 0.6111 & \textbf{0.6859} & 0.7143 & 0.4047 & \textbf{0.5437} & 0.4167\\
\hline
\end{tabular}
\begin{tablenotes}
\item[1]{We use bold font to highlight the rAPFD values with better results compared to Table~\ref{table:DDP} and the optimal value 1.}
\end{tablenotes}
\end{threeparttable}
\vspace{-0.4cm}
\end{table*}

% \vspace{-0.05cm}
\emph{Hypothesis:} The main difference between the two kinds of subjects lies in the failure rate, which causes data imbalance problem. \textbf{RL-based techniques usually need a large amount of data to be effective~\cite{hester2018deep}.} However, on the less-failure subjects, there are too few failing tests for the reinforcement learning model to learn a mature strategy. 
\textbf{SL-based techniques have different performance when facing imbalanced data. However, some ensemble models~\cite{lopez2013insight} have the advantage to deal with imbalanced data.} MART is a powerful ensemble model which continuously constructs weak classifiers to minimize the error.
% \dan{what does ``minimize the residual'' mean?}
Therefore, MART performs the best on less-failure subjects.

% \vspace{-0.3cm}
% \vspace{-0.05cm}
\emph{Validation:} To validate the preceding hypothesis, we design an additional experiment. We use the over-sampling approach SMOTE to synthesize positive samples to alleviate data imbalance. In particular, we apply SMOTE~\cite{chawla2002smote} to the training data of the 11 subjects. Then we train the studied TCP techniques using synthesized data and test them on the original testing data. Table~\ref{table:DDP-smote} illustrates the rAPFD results on the testing data. From the table, SMOTE boosts both SL-based techniques (MART, RankNet, RankBoost, L-MART) and RL-based techniques (ACER-PA). Especially on less-failure subjects, more techniques produce the optimal sequence in each cycle, as more ``1'' rAPFD values appear. These results show that imbalanced data indeed hamper the effectiveness of TCP techniques, but some data pre-processing approaches like SMOTE can release the power of existing TCP techniques. 

\begin{center}
\fcolorbox{black}{gray!10}{\parbox{0.95\linewidth}{
\textbf{Finding 1:} On more-failure subjects and less-failure subjects, the performance of ML-based TCP techniques is different. On more-failure subjects, the RL-based technique ACER-PA performs the best, while on less-failure subjects, the SL-based technique MART performs the best. This phenomenon is caused by data imbalance. 
% Overall, MART achieves the best performance on all subjects.

\textbf{Actionable suggestions:} Techniques dealing with data imbalance well may perform better for TCP in CI, therefore, cost-sensitive, ensemble algorithms or other robust techniques are promising in this scenario. Besides, pre-processing approaches alleviating data imbalance like SMOTE can boost the existing TCP techniques.
}}
\end{center}
% \vspace{-0.1cm}

\subsubsection{RQ1.2: Effectiveness across CI cycles}

% Frequent code change in CI results in many CI cycles, each of which may triggers regression testing.
% In each CI cycle, a TCP technique schedules the test suite based on the code features and the execution history of tests. Recall that the two categories of TCP techniques have different learning strategies.

The two categories of TCP techniques have different learning strategies.
% For example, the supervised-learning based techniques may take the execution information as model input, while the reinforcement-learning based techniques may use the actual execution feedback of test cases (i.e., passing or failing). 
While the SL-based techniques use only instances in early cycles as training data, the RL-based techniques refine their models continuously in coming cycles. In other words, SL-based models are fixed once they finish training with data from early cycles, but RL-based models may improve their effectiveness if more CI cycle data are available. Therefore, we investigate the effectiveness of the studied TCP techniques across CI cycles.

% \vspace{-1cm}
In particular, we plot the trend of rAPFD values for the compared TCP techniques across CI cycles in Fig.~\ref{fig-ci-cycle}, where the x-axis represents the number of failing CI cycles ordered by timeline. We report the average rAPFD at every 5 commits for readability. Due to space limitation, we only present results on more-failure subjects \textit{jedis}, \textit{nfe} and \textit{spring-data-redis}, because they have more failing cycles, which presents a clearer trend. Results on other subjects are referred to our website. 

From Fig.~\ref{fig-ci-cycle}, the performance of the techniques is
more stable on \textit{nfe}, but tends to fluctuate on \textit{jedis} and \textit{spring-data-redis}.
In particular, for RL-based techniques, the rAPFD tends to increase in some cases (e.g., ACER-PA and PPO2-PO on \textit{jedis}). For SL-based techniques, the rAPFD tends to decrease in some cases (e.g., MART and L-MART on \textit{jedis}). Why do the TCP techniques perform differently across CI cycles?

\emph{Hypothesis:} \textbf{Along with software evolution, the performance of the TCP techniques changes across CI cycles, because the increasing number of CI cycles brings two side-effects: more code change (including subject code evolution and test removal/addition) and more training data.} We hypothesize that the two side-effects make techniques perform differently. 
\textbf{First, more code change indicates that the previous prioritization strategy may not be suitable in later CI cycles.} For SL-based TCP techniques, the potential big difference in code and tests across many CI cycles may hamper their effectiveness, because these techniques regard tests of early CI cycles as training data and cannot adapt to the dynamic environment afterward. However, for RL-based techniques, they have the ability to adapt to the highly variable environment~\cite{bertolino2020learning}. Therefore, the first side-effect affects SL-based techniques to a larger extent. \textbf{Second, more training data may boost RL-based techniques, because they continuously get trained using the coming data, while SL-based techniques cannot.} Therefore, the second side-effect affects RL-based techniques to a larger extent. 
To sum up, we hypothesize the two side-effects to be the reason why techniques perform differently on earlier and later CI cycles.

% \begin{table}[tp]
% \vspace{-0.2cm}
% \footnotesize
% \centering
% \setlength{\abovecaptionskip}{0cm} 
% % \setlength{\belowcaptionskip}{-0.2cm}
% \caption{Average training and prediction time (in seconds)}
% % \vspace{-0.3cm}
% \label{tab:time}
% \begin{tabular}{r|c|c}
% \hline \text { } & \text{training time}& \text{prediction time}\\
% \hline \text { MART (SL,PA)} & 0.0826-0.9157 & 0.0062-0.0340\\
% \hline \text { RankNet (SL,PA)} & 0.0188-1.2222 & 0.0019-0.0111\\
% \hline \text { RankBoost (SL,PA)} & 0.0446-0.6088 & 0.0019-0.0127\\
% \hline \text { CA (SL,LI)} & 0.0493-0.5110 & 0.0017-0.0111\\
% \hline \text { L-MART (SL,PA)} & 0.0056-0.0633 & 0.0021-0.0127\\
% \hline \text { DeepOrder (SL,PO)} & 0.0000-0.0500 & 0.0191-0.0241\\
% \hline \text { RL (RL,PO)} & 0.8018-2.1837 & 0.0011-0.0024\\
% % \hline \text { RL-MLP } & 71.582-849.780 & 0.051-0.646\\
% \hline \text { COLEMAN (RL,PO)} & 0.0000-0.0030 & 0.0001-0.0030\\
% \hline \text { PPO2-PO (RL,PO)} & 65.5094-285.4404 & 0.6155-0.7852\\
% \hline \text { ACER-PA (RL,PA)} & 56.4518-219.4478 & 1.5157-2.3241\\
% \hline \text { PPO1-LI (RL,LI)} & 59.6611-1093.5445 & 0.9940-308.3892\\
% \hline
% \end{tabular}
% \vspace{-0.6cm}
% \end{table}

\begin{table*}[!htp]
\centering
\footnotesize
\setlength{\abovecaptionskip}{0cm} 
\setlength{\belowcaptionskip}{0cm}
\begin{threeparttable}
\caption{Average rAPFD for pretrained TCP techniques}
% \vspace{-0.3cm}
\label{table:DDP-finetune}
\begin{tabular}{r|cccc|cccccc}
\hline \multirow{2}{*}{Subjects} & \multicolumn{4}{c|}{More-failure subjects} & \multicolumn{6}{c}{Less-failure subjects} \\
\cline{2-11}  & \text{bcel} & \text{jedis} & \text{jsprit} & \text{spring-data-redis} &\text{csv} &\text{dbcp} & \text{text} & \text{java-faker} & \text{jsoup} & \text{maxwell}\\
\hline \text { MART (SL,PA) } & 0.6351 & \textbf{1.0000} & \textbf{1.0000} & 0.8595 & \textbf{1.0000} & \textbf{1.0000} & \textbf{1.0000} & \textbf{1.0000} & \textbf{1.0000} & \textbf{1.0000}\\
\hline \text { RankNet (SL,PA) } & \textbf{1.0000} & \textbf{0.9358} & \textbf{1.0000} & \textbf{0.6677} & \textbf{1.0000} & \textbf{1.0000} & \textbf{1.0000} & \textbf{0.9904} & \textbf{1.0000} & \textbf{1.0000} \\
\hline \text { RankBoost (SL,PA) } & \textbf{1.0000} & \textbf{1.0000} & 0.3801 & 0.4310 & \textbf{0.9722} & \textbf{0.6266} & \textbf{1.0000} & 0.9535 & 0.3754 & 0.5000 \\
\hline \text { CA (SL,LI) } & 0.5893 & \textbf{0.9812} & \textbf{1.0000} & 0.7267 & \textbf{0.9722} & \textbf{0.5559} & \textbf{0.5714} & \textbf{0.9417} & 0.7943 & \textbf{0.8750} \\
\hline \text { L-MART (SL,PA) } & 0.5113 & \textbf{0.5870} & \textbf{0.9741} & 0.3981 & 0.1667 & \textbf{0.9161} & \textbf{0.7143} & \textbf{0.8546} & 0.2781 & \textbf{0.9792} \\
\hline \text { DeepOrder (SL,PO) } & \textbf{1.0000} & 0.5635 & \textbf{0.8594} & \textbf{0.8802} & \textbf{0.8333} & 0.6447 & 0.7143 & \textbf{0.6499} & 0.6832 & 0.9460 \\
\hline \text { RL (RL,PO) } & 0.4349 & 0.4176 & 0.4977 & \textbf{0.4885} & 0.5278 & 0.1332 & \textbf{0.8571} & 0.4414 & \textbf{0.7189} & 0.6098 \\
% \hline \text { RL-MLP } & 0.4578 & 0.5468 & 0.4943 & 0.4797 & 0.5015 & 0.1795 & 0.3997 & 0.2500 & 0.4877 & 0.2105 & 0.7626\\
\hline \text { COLEMAN (RL,PO) } & 0.8768 & 0.8031 & \textbf{0.0826} & 0.3829 & 0.9722 & 0.8092 & 0.1429 & \textbf{0.5584} & 0.6623 & 0.3703 \\
\hline \text { PPO2-PO (RL,PO) } & 0.6537 & 0.6370 & \textbf{0.9157} & 0.7146 & 0.2500 & \textbf{0.3240} & \textbf{0.8571} & \textbf{0.3174} & \textbf{0.4902} & 0.0312 \\
\hline \text { ACER-PA (RL,PA) } & 0.7349 & 0.7891 & 0.5255 & 0.4911 & \textbf{0.6389} & \textbf{0.3865} & 0.2857 & \textbf{0.4072} & \textbf{0.5300} & \textbf{0.4223} \\
\hline \text { PPO1-LI (RL,LI) } & \textbf{0.4404} & 0.4597 & \textbf{0.7943} & 0.4728 & 0.5000 & \textbf{0.6513} & 0.8571 & 0.5763 & 0.3993 & 0.1894 \\
\hline
\end{tabular}
\begin{tablenotes}
\item[1]{We use bold font to highlight the rAPFD values with better results compared to Table~\ref{table:DDP} and the optimal value 1.}
\end{tablenotes}
\end{threeparttable}
\vspace{-0.6cm}
\end{table*}

\emph{Validation:} 
To validate the preceding hypothesis, we design two additional experiments to investigate the impact of the two side-effects. 

First, to investigate the impact of more training data, we use data from other subjects to ``pretrain'' the techniques, then we ``finetune'' the techniques on their original training instances and test them on the original testing instances. 
To be more specific, on each subject, we first use data from \textit{nfe} to train the techniques (as \textit{nfe} has the most instances) and get the pretrained versions of them. Then we finetune and test the pretrained techniques on the original dataset. We run the 11 pretrained TCP techniques on the 10 subjects (excluding \textit{nfe} because it is used to pretrain the techniques). Table~\ref{table:DDP-finetune} presents the average rAPFD results for the pretrained techniques. From the table, we observe that pretraining can greatly improve the performance of both RL-based techniques and SL-based techniques. \textbf{Specifically, the pretrained MART and RankNet produce the optimal sequence for 80\% and 70\% subjects respectively, while for the original versions, they only produce the optimal sequence for 50\% and 10\% subjects respectively (on the ten subjects excluding \textit{nfe}). Besides, the pretrained PPO2-PO and ACER-PA perform better on 67\% and 83\% less-failure subjects respectively. It is also worth noting that the pretrained MART achieves state-of-the-art TCP effectiveness.} While producing the optimal sequence on 80\% subjects, it also have stable and satisfactory performance on the rest 20\% subjects. According to the experimental results, we can conclude that training data play an important role in affecting the effectiveness of TCP techniques.

\begin{figure}
  \centering
  % \vspace{cm}
  \includegraphics[width=0.4\textwidth]{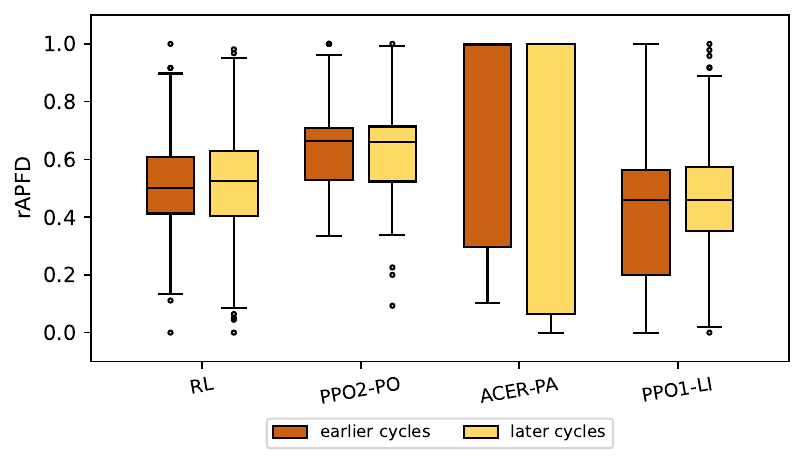} 
  \vspace{-0.5cm}
  \caption{rAPFD Results in earlier cycles and later cycles}
  \vspace{-0.4cm}
  \label{fig-total-freeze} 
\end{figure}

Second, to investigate the impact of more code change, we stop training the RL-based techniques once they get trained on the first 2000 instances (i.e., we train them with the same amount of data as SL-based techniques) and thus get their ``early-stopped'' versions. To be more specific, as RL-based techniques can continuously refine their strategies based on coming data and adapt to the code change, their ``early-stopped''
% \dan{add a footnote to explain early-stop}
versions have not adapted to the code change in the later cycles. Therefore, we investigate the impact of code change by comparing the performance of the early-stopped RL-based techniques in the earlier and later cycles. We run four RL-based techniques (RL, PPO2-PO, ACER-PA, PPO1-LI) on the three more-failure subjects (i.e., \textit{jedis}, \textit{nfe}, \textit{spring-data-redis}). For earlier and later cycles, we use the first 30 and last 30 cycles in the testing cycles respectively. Fig.~\ref{fig-total-freeze} presents the results in the earlier and later cycles. 
From the figure, the early-stopped techniques perform almost the same in the earlier and later cycles, indicating that the models only trained on earlier cycles do not have lower effectiveness on later cycles. 
According to the experimental results, during the certain period of time, we can conclude that code change does not affect the effectiveness of TCP techniques to a large extent in the CI scenario. The conclusion is reached on ML-based techniques, but is consistent with previous study~\cite{DBLP:conf/icse/LuLCZHZ016} on traditional and time-aware TCP techniques.

% \vspace{-0.1cm}
\begin{center}
\fcolorbox{black}{gray!10}{\parbox{0.95\linewidth}{
\textbf{Finding 2:} The performance of the TCP techniques changes across CI cycles mainly caused by the changing amount of training data, rather than code evolution and test removal/addition. The RL-based technique may perform well on later CI cycles, because they continuously get trained using the coming data, while SL-based techniques cannot. Therefore, more training data lead to better effectiveness for TCP in CI.

\textbf{Actionable suggestions:} Cross-subject data can be used to pretrain TCP techniques. With more training data available, TCP techniques get fully trained and perform better.
}}
\end{center}

\begin{table}[tp]
\vspace{-0.2cm}
\footnotesize
\centering
\setlength{\abovecaptionskip}{0cm} 
\caption{Average training and prediction time (in seconds)}
% \vspace{-0.3cm}
\label{tab:time}
\begin{tabular}{r|c|c}
\hline \text { } & \text{training time}& \text{prediction time}\\
\hline \text { MART (SL,PA)} & 0.0826-0.9157 & 0.0062-0.0340\\
\hline \text { RankNet (SL,PA)} & 0.0188-1.2222 & 0.0019-0.0111\\
\hline \text { RankBoost (SL,PA)} & 0.0446-0.6088 & 0.0019-0.0127\\
\hline \text { CA (SL,LI)} & 0.0493-0.5110 & 0.0017-0.0111\\
\hline \text { L-MART (SL,PA)} & 0.0056-0.0633 & 0.0021-0.0127\\
\hline \text { DeepOrder (SL,PO)} & 0.0000-0.0500 & 0.0191-0.0241\\
\hline \text { RL (RL,PO)} & 0.8018-2.1837 & 0.0011-0.0024\\
% \hline \text { RL-MLP } & 71.582-849.780 & 0.051-0.646\\
\hline \text { COLEMAN (RL,PO)} & 0.0000-0.0030 & 0.0001-0.0030\\
\hline \text { PPO2-PO (RL,PO)} & 65.5094-285.4404 & 0.6155-0.7852\\
\hline \text { ACER-PA (RL,PA)} & 56.4518-219.4478 & 1.5157-2.3241\\
\hline \text { PPO1-LI (RL,LI)} & 59.6611-1093.5445 & 0.9940-308.3892\\
\hline
\end{tabular}
\vspace{-0.6cm}
\end{table}

% \vspace{-0.2cm}
\subsection{RQ2: Efficiency Comparison}
\label{sec:rq2}
% \vspace{-0.1cm}

% A large overhead makes TCP techniques unnecessary because TCP aims to accelerate fault detection. 
% As commits come frequently, the time between commits, i.e., time left for training and prediction is relatively short. 
% We investigate the efficiency of the ML-based TCP techniques in terms of training and prediction time. 

Table~\ref{tab:time} presents the average training and prediction time in each cycle for all subjects. Due to space limitation, we give the time range of all subjects and more detailed data are referred to our website. Note that the average training time of SL-based techniques refers to their total training time divided by the number of training CI cycles.
% Note that a SL-based technique is trained only once, while a RL-based technique is continuously trained in each CI cycle.
From this table, the RL-based techniques (e.g., RL, PPO2-PO, ACER-PA, and PPO1-LI) generally have much longer training time, especially for PPO2-PO, ACER-PA, and PPO1-LI. It is because for each cycle, the three techniques are trained with 200 episodes, while in each episode, they are trained at the minimum of $n*log_2(n)$ steps, where $n$ represents the number of tests in the cycle, to ensure enough training~\cite{bagherzadeh2021reinforcement}. Therefore, although ACER-PA achieves pretty good TCP effectiveness (according to Section~\ref{sec:rq1}), it has longer training time. The large number of episodes and comparisons in training make it among the least efficient. 

Prediction time is more important than training time for ML-based TCP techniques, because training can be conducted offline but prediction has to be conducted online. According to Table~\ref{tab:time}, the prediction time of ACER-PA and PPO1-LI is much longer than other techniques, indicating that they are the least efficient. Especially for PPO1-LI, its average prediction time for \textit{nfe} is 308.3892 seconds, which is extremely long. However, the prediction time of other TCP techniques is less than 1 second, which is ignorable. 

% \vspace{-0.1cm}
\begin{center}
\fcolorbox{black}{gray!10}{\parbox{0.95\linewidth}{
\textbf{Finding 3:} RL-based techniques generally have much longer training time than SL-based techniques. In particular, ACER-PA is among the least efficient techniques although achieves high effectiveness, while SL-based techniques are all relatively efficient.
}}
\end{center}
% \vspace{-0.4cm}

\begin{table*}[!htp]
\centering
\footnotesize
\setlength{\abovecaptionskip}{0cm} 
\setlength{\belowcaptionskip}{-0.2cm}
\caption{Average NTR for compared TCP techniques}
% \vspace{-0.3cm}
\label{table:ntr}
\begin{tabular}{r|ccccc|cccccc}
\hline \multirow{2}{*}{Subjects} & \multicolumn{5}{c|}{More-failure subjects} & \multicolumn{6}{c}{Less-failure subjects} \\
\cline{2-12}  & \text{bcel} & \text{jedis} & \text{jsprit} & \text{nfe} & \text{spring-data-redis} &\text{csv} &\text{dbcp} & \text{text} & \text{java-faker} & \text{jsoup} & \text{maxwell}\\
\hline \text { MART (SL,PA) } & 0.6970 & 0.5029 & \textbf{0.9991} & \textbf{0.9969} & \textbf{0.9980} & \textbf{0.9994} & \textbf{0.9517} & \textbf{0.9772} & \textbf{0.9907} & \textbf{0.9845} & \textbf{0.8615}\\
\hline \text { RankNet (SL,PA) } & 0.6108 & 0.8607 & 0.9848 & 0.9785 & 0.9782 & \textbf{0.9994} & 0.2517 & 0.9769 & 0.7721 & 0.1189 & 0.3411 \\
\hline \text { RankBoost (SL,PA) } & \textbf{0.6977} & 0.7262 & \textbf{0.9991} & 0.9829 & 0.9954 & 0.9993 & 0.0737 & 0.9525 & 0.9874 & 0.9753 & 0.7316\\
\hline \text { CA (SL,LI) } & 0.2516 & 0.4990 & 0.2673 & 0.8919 & 0.9925 & 0.5669 & 0.3477 & 0.0731 & 0.1767 & 0.1042 & 0.5271 \\
\hline \text { L-MART (SL,PA) } & 0.6974 & 0.7102 & 0.8797 & 0.9754 & 0.9672 & 0.3784 & 0.0200 & 0.8888 & 0.4832 & 0.9409 & 0.7721 \\
\hline \text { DeepOrder (SL,PO) } & 0.4895 & 0.8643 & 0.0704 & 0.6394 & 0.0144 & 0.0001 & 0.0810 & 0.0246 & 0.1299 & 0.0441 & 0.2707 \\
\hline \text { RL (RL,PO) } & 0.3083 & 0.4738 & 0.3846 & 0.9232 & 0.7870 & 0.6452 & 0.7165 & 0.0820 & 0.7974 & 0.0346 & 0.6204 \\
% \hline \text { RL-MLP } & 0.4578 & 0.5468 & 0.4943 & 0.4797 & 0.5015 & 0.1795 & 0.3997 & 0.2500 & 0.4877 & 0.2105 & 0.7626\\
\hline \text { COLEMAN (RL,PO) } & 0.5039 & 0.5631 & 0.4177 & 0.9225 & 0.9315 & 0.9991 & 0.8587 & 0.0003 & 0.7641 & 0.8886 & 0.2459 \\
\hline \text { PPO2-PO (RL,PO) } & \textbf{0.6977} & \textbf{0.8802} & 0.9437 & 0.9410 & 0.9951 & 0.9991 & 0.4749 & 0.9769 & 0.7272 & 0.9276 & 0.5515 \\
\hline \text { ACER-PA (RL,PA) } & \textbf{0.6977} & 0.8497 & 0.9981 & 0.9837 & 0.9949 & 0.9960 & 0.8707 & 0.9525 & 0.8034 & 0.9344 & 0.5500 \\
\hline \text { PPO1-LI (RL,LI) } & 0.2631 & 0.4950 & 0.6972 & 0.9210 & 0.7900 & 0.6901 & 0.7254 & \textbf{0.9772} & 0.5222 & 0.7512 & 0.8526 \\
\hline
\end{tabular}
\vspace{-0.5cm}
\end{table*}

% \vspace{-0.3cm}
\subsection{RQ3. Applicability}
% \vspace{-0.1cm}

We investigate the applicability of the TCP techniques from three aspects: First, we compare the commit interval with the training and prediction time of the techniques.
% Although training is conducted offline, 
It should be guaranteed that before the next commit, the techniques have completed training
% , especially for RL-based techniques that train after each cycle, 
and prediction. Otherwise, the adapted model is not ready and the test suite is not scheduled when the new cycle starts. Second, we compare the test duration with the prediction time of the techniques. If the prediction time is longer than the total test execution time, TCP becomes unnecessary. Third, we investigate to what extent can the TCP techniques save time. In each cycle, once the first failure is revealed, the developers are reminded that bugs exist and can start to fix them. We use the indicator NTR introduced in Section~\ref{sec:metric_selection} to measure the time reduction ratio.

% and Table~\ref{tab:sub}
Table~\ref{tab:sub} presents the average commit interval for each subject, which ranges from 11234.82 seconds to 25343.38 seconds. As Table~\ref{tab:time} shows, the training time of most techniques is shorter than 3 seconds, which is negligible to the commit interval. For the least efficient techniques PPO2-PO, ACER-PA, and PPO1-LI, the training time is shorter than 1100 seconds. Moreover, the prediction time for each technique is shorter than the training time. Therefore, on average, all techniques have sufficient time for training and prediction. We provide the ratio of commits in each project for which training and prediction is possible on our website.

Table~\ref{tab:sub} presents the average test duration for each subject, which ranges from 0.065 to 2123.381 seconds. For some subjects, e.g., \textit{jedis}, \textit{jsprit}, and \textit{bcel}, executing all the tests only costs less than 33 seconds after selection. However, PPO1-LI costs 121.858 seconds, 122.823 seconds, and 102.9197 seconds respectively to predict the results for the three subjects. Therefore, PPO1-LI may be inapplicable in some subjects because developers have to wait more time for prediction than simply running all the tests. Other techniques all have relatively short prediction time but may be unnecessary in some cases, e.g., a cycle costs 0.65 seconds to execute all selected tests in \textit{bcel}. Therefore, we suggest developers applying TCP techniques only if the estimated test duration is longer than a predefined threshold (the execution time of a test can be estimated by averaging its previous execution time~\cite{spieker2017reinforcement}).

Table~\ref{table:ntr} shows the average NTR for the techniques in each subject. We highlight the best technique for each subject in bold font. From the table, MART is the best technique achieving the best results on 9 out of 11 subjects. Moreover, most of its NTR values are above 0.95, indicating that the time spent to reveal the first failure costs less than 5\% of the total duration. 
% Therefore, developers can get failure results sooner and start to fix the bugs earlier. 
Besides MART, ACER-PA is also effective and most of its NTR values are above 0.8. 
The results are consistent with our previous analysis in Section~\ref{sec:rq1.1}.
Other techniques may perform very poorly on one subject and the unstable performance makes them less applicable. For example, the NTR value of DeepOrder on \textit{csv} is 0.0001, indicating that the first failure is detected at the end. Note that Table~\ref{table:ntr} and Table~\ref{table:DDP} yield similar results in terms of the best technique in each subject. Some difference exists because NTR only considers the rank of the first failing test, but rAPFD considers the whole picture, i.e., the ranks of all failing tests. 

% \vspace{-0.2cm}
\begin{center}
\fcolorbox{black}{gray!10}{\parbox{0.95\linewidth}{
\textbf{Finding 4:} The average training and prediction time of all techniques is shorter than the average commit intervals in all subjects. However, PPO1-LI's prediction time exceeds the total test execution time in some subjects, which makes it inapplicable. Except for PPO1-LI, other techniques are generally applicable to the CI context. A good TCP technique (e.g., MART) can help developers save more than 95\% time cost in most subjects.

\textbf{Actionable suggestions}: Test duration should be estimated to assess the necessity of applying any TCP techniques. The overhead of TCP techniques deserves more attention and may become the bottleneck of application in practice. 
}}
\end{center}

% \begin{center}
% \fcolorbox{black}{gray!10}{\parbox{0.95\linewidth}{
% \textbf{Actionable suggestions}: Test duration should be estimated to assess the necessity of applying any TCP techniques. The overhead of TCP techniques deserves more attention and may become the bottleneck of application in practice. 
% }}
% \end{center}

% \vspace{-0.3cm}

% \vspace{-0.3cm}
\section{Threats to validity}
\label{sec:valid}
% \vspace{-0.2cm}

\textbf{Threats to internal validity} mainly come from possible errors in implementing the compared techniques. To mitigate the threats, we use the mature library \textit{Ranklib} and the released code to implement the techniques
~\cite{bagherzadeh2021reinforcement,bertolino2020learning,do2020multi}. 
Besides, we strictly follow the description and use the published processing scripts to extract the features that each technique takes in.
% Besides, we use the same parameter settings as the previous work~\cite{bagherzadeh2021reinforcement,bertolino2020learning,do2020multi,sharif2021deeporder}. 

\textbf{Threats to external validity} mainly come from subjects, including the commits and tests used in the study, and the potential flaky tests. As the existing datasets do not contain sufficient feature information of studied techniques, we select 11 popular and large-scale projects from GitHub, which are representative in terms of their project scale, popularity, numbers of CI cycles and tests. The number and scale of subjects used in our study are comparable to prior empirical studies~\cite{bertolino2020learning, bagherzadeh2021reinforcement,DBLP:conf/icse/LuLCZHZ016,haghighatkhah2018test,yang2020systematic,hemmati2017prioritizing}. Moreover, following previous work~\cite{bertolino2020learning}, we select only test classes covering changed files as candidate tests, which may filter out flaky tests. Because flaky test failures are usually introduced into the program before the latest version~\cite{luo2014empirical}.

\textbf{Threats to construct validity} come from the metrics we use and the recognition of more-failure and less-failure subjects. To reduce the former threats, we use the existing metric NTR to measure the applicability of the techniques~\cite{do2020multi}. For the effectiveness metric, we first analyze the existing metrics systematically. Then we take advantage of the widely-used metric APFD~\cite{bagherzadeh2021reinforcement,sharif2021deeporder} and do min-max normalization to fit it in the CI context. The normalization does not change the monotonicity of APFD. We may use more appropriate new metrics in our future work. To reduce the latter threats, we use the 1\% failure rate as the boundary because He et al.~\cite{he2009learning} reported three levels of data imbalance: the ratios of 100:1, 1000:1, and 10,000:1 between majority and minority classes.
\section{Discussion}
% \vspace{-0.2cm}
% \dan{fill in the description.}
% In practice, CI occurs frequently, resulting in frequent rebuilding and testing.
To alleviate the resource consumption in CI, two orthogonal directions in CI have been explored in the literature, build selection/prioritization and test selection/prioritization~\cite{jin2021helped}. The former targets prioritizing or selecting builds to detect failure early, whereas the latter focuses on tests instead of builds. 
% In particular, Beller et al.~\cite{beller2017oops} pointed out that in their studied 1,108 Java projects on Travis CI, the ratio of commits with at least one failed test case is 10.3\% on average.
To reduce the building and the following testing time cost, Liang et. al~\cite{liang2018redefining} proposed prioritizing commits based on test execution history, and Jin~\cite{jin2021reducing} proposed models to predict commits and builds that can be skipped. 
% In this paper we present the first extensive study on ML-based TCP techniques in CI by controlling these factors (e.g., commit prioritization~\cite{jin2021helped} and parallel testing~\cite{zhou2021parallel}), 
In this paper we present the first extensive study on ML-based TCP techniques in CI by controlling these factors (e.g., build prioritization and selection), 
and plan to investigate the combination of the two directions in the future. 

%\yifan{About our selected subjects, one may find that the number of failure builds is rather small. Beller et al.~\cite{beller2017oops} showed that for all 1,108 Java projects with test executions on Travis CI, the ratio of builds with at least one failed test case has a median of 2.9\% and a mean of 10.3\%. However, this does not influence the necessity of TCP because TCP focuses on detecting fault eariler on failure builds. For those potentially passing builds, e.g., builds caused by cosmetic commits, developers may benefit from the CI skipping techniques~\cite{jin2021reducing,jin2022hybridcisave}, which target at an orthogonal domain to TCP. Besides, in real world, developers may leverage more computing resources to do parallel test to get the time down. TCP techniques can also be applied in parallel scenario~\cite{zhou2021parallel} although need some adaption. We leave this problem to our future work.}

% As the only major comment, I think the paper would have greatly improved with a nice discussion on when to use each algorithm, depending on several factors such as software system size, CI cycle frequency, average commit time, mean failure ratio, etc. While MART seems to be the most appropriate algorithm in most cases, the results also show that other algorithms can be better in certain scenarios (e.g., DeepOrder).

Among existing studies, only the works conducted by Bertolino et al.~\cite{bertolino2020learning} and Bagherzadeh et al.~\cite{bagherzadeh2021reinforcement} are close to this paper (i.e., comparing RL-based and SL-based techniques). In particular, this paper confirms the following findings (in our Finding 1, Finding 3) by including more advanced techniques: (1) Pairwise and ensemble learning algorithms have better effectiveness. (2) RL algorithms require long training time. This paper extends the existing works by including the following new findings (in our Finding 1, Finding 2, Finding 4): (1) ML-based TCP techniques perform differently on different subjects due to failure rate. (2) The performance change across CI cycles is mainly caused by more training data rather than code evolution. (3) Most TCP techniques are generally applicable to CI. Moreover, this paper also produces findings (in our Finding 1, Finding 4) that contradicted existing studies, i.e., PPO1-LI is inapplicable in some subjects and ACER-PA does not perform better than MART on less-failure subjects.

We here provide a discussion outlining the recommended use scenarios of the TCP techniques based on our experimental results. When facing more-failure subjects, we suggest using ACER-PA as the TCP technique. Conversely, when facing less-failure subjects, we suggest using MART. This recommendation is based on the fact that ACER-PA and MART have demonstrated superior performance on these respective types of subjects in our experiment. To estimate the failure ratio, one can rely on the test results obtained during early CI cycles. If a subject has a high CI cycle frequency and, consequently, a low average commit interval, we suggest using SL-based techniques such as MART. This is because RL-based techniques tend to be more time-consuming for training and predicting. If a subject has a relatively long evolution process, we suggest using RL-based techniques like ACER-PA. RL-based techniques have the advantage of continuously improving their strategies as new data become available, without requiring retraining after a certain period, which is in contrast to SL-based techniques~\cite{yaraghi2022scalable}. If the subject is newly-created and has few commits which cannot support sufficient training, or the developers want to improve the performance of TCP techniques, we suggest collecting test execution data from other high-quality subjects. These data can then be used to train the ML models and improve their effectiveness.

% \vspace{-0.2cm}
\section{Conclusion}
% \vspace{-0.1cm}
%Test case prioritization in CI has gained researchers' attention and many ML-based prioritization techniques have been proposed recently. Although some previous work claim that RL-based techniques may be better than SL-based techniques, it lacks a comprehensive study to confirm this assumption. 
To learn how ML-based TCP techniques perform in CI, we present the first comprehensive study on 11 GitHub projects, including 11 state-of-the-art ML-based TCP techniques.  
% According to the study, we have several interesting findings and actionable suggestions. 
In our study, we systematically analyze the effectiveness, efficiency, and applicability of existing TCP techniques and get a series of findings and actionable suggestions. Our study gives a comprehensive view of TCP in CI as well as future directions for improving TCP techniques in CI.
% \section{Data Availability}
% The replication package is available at \url{https://zenodo.org/badge/latestdoi/529202344}. We provide the dataset, code, and result for replicating our experiments.

\section*{Acknowledgment}

This work was supported by National Natural Science Foundation of China under Grant No. 62232001 and No. 62232003.

\bibliographystyle{IEEEtran}
\bibliography{ref.bib}
\end{document}